\documentclass[manuscript,screen]{acmart}

\AtBeginDocument{%
  \providecommand\BibTeX{{%
    \normalfont B\kern-0.5em{\scshape i\kern-0.25em b}\kern-0.8em\TeX}}}

\setcopyright{acmcopyright}
\copyrightyear{2018}
\acmYear{2018}
\acmDOI{XXXXXXX.XXXXXXX}

\acmConference[Conference acronym 'XX]{Make sure to enter the correct
  conference title from your rights confirmation emai}{June 03--05,
  2018}{Woodstock, NY}
\acmPrice{15.00}
\acmISBN{978-1-4503-XXXX-X/18/06}

\begin{document}

\title{Recipe Generation from Unsegmented Cooking Videos}

\author{Taichi Nishimura}
\email{taichitary@gmail.com}
\authornotemark[1]
\affiliation{%
  \institution{Graduate School of Informatics, Kyoto University}
  \streetaddress{Yoshidahonmachi}
  \city{Sakyo-ku}
  \state{Kyoto}
  \country{Japan}
  \postcode{606--8501}
}

\author{Atsushi Hashimoto}
\email{atsushi.hashimoto@sinicx.com}
\authornotemark[2]
\affiliation{%
  \institution{OMRON SINIC X Corporation}
  \streetaddress{5--24--5 Hongo}
  \city{Bunkyo-ku}
  \state{Tokyo}
  \country{Japan}
  \postcode{113--0033}
}

\author{Yoshitaka Ushiku}
\email{yoshitaka.ushiku@sinicx.com}
\authornotemark[2]
\affiliation{%
  \institution{OMRON SINIC X Corporation}
  \streetaddress{5--24--5 Hongo}
  \city{Bunkyo-ku}
  \state{Tokyo}
  \country{Japan}
  \postcode{113--0033}
}

\author{Hirotaka Kameko}
\email{kameko@i.kyoto-u.ac.jp}
\authornotemark[3]
\affiliation{%
  \institution{Academic Center for Computing and Media Studies, Kyoto University}
  \streetaddress{Yoshidahonmachi}
  \city{Sakyo-ku}
  \state{Kyoto}
  \country{Japan}
  \postcode{606--8501}
}

\author{Shinsuke Mori}
\email{forest@i.kyoto-u.ac.jp}
\authornotemark[3]
\affiliation{%
  \institution{Academic Center for Computing and Media Studies, Kyoto University}
  \streetaddress{Yoshidahonmachi}
  \city{Sakyo-ku}
  \state{Kyoto}
  \country{Japan}
  \postcode{606--8501}
}

\renewcommand{\shortauthors}{Taichi Nishimura, et al.}

\begin{abstract}
This paper tackles recipe generation from unsegmented cooking videos, a task that requires agents to (1) extract key events in completing the dish and (2) generate sentences for the extracted events.
Our task is similar to dense video captioning (DVC), which aims at detecting events thoroughly and generating sentences for them. However, unlike DVC, in recipe generation, recipe story awareness is crucial, and a model should extract an appropriate number of events in the correct order and generate accurate sentences based on them.
We analyze the output of the DVC model and confirm that although (1) several events are adoptable as a recipe story, (2) the generated sentences for such events are not grounded in the visual content.
Based on this, we set our goal to obtain correct recipes by selecting oracle events from the output events and re-generating sentences for them.
To achieve this, we propose a transformer-based multimodal recurrent approach of training an event selector and sentence generator for selecting oracle events from the DVC's events and generating sentences for them.
In addition, we extend the model by including ingredients to generate more accurate recipes. The experimental results show that the proposed method outperforms state-of-the-art DVC models. We also confirm that, by modeling the recipe in a story-aware manner, the proposed model outputs the appropriate number of events in the correct order.
\end{abstract}



\def\tabref#1{Table \ref{#1}}
\def\figref#1{Fig. \ref{#1}}
\def\secref#1{Section \ref{#1}}
\def\eqref#1{Eq (\ref{#1})}
\newcommand{\etal}{\textit{et al}.~}

\newcommand{\x}{\mbox{\boldmath $x$}}
\newcommand{\y}{\mbox{\boldmath $y$}}
\newcommand{\h}{\mbox{\boldmath $h$}}
\newcommand{\g}{\mbox{\boldmath $g$}}
\newcommand{\w}{\mbox{\boldmath $w$}}
\newcommand{\ei}{\mbox{\boldmath $e$}}
\newcommand{\ui}{\mbox{\boldmath $u$}}
\newcommand{\ai}{\mbox{\boldmath $a$}}

\newcommand{\X}{\mbox{\boldmath $X$}}
\newcommand{\A}{\mbox{\boldmath $A$}}
\newcommand{\Y}{\mbox{\boldmath $Y$}}
\newcommand{\U}{\mbox{\boldmath $U$}}
\newcommand{\C}{\mbox{\boldmath $C$}}
\newcommand{\E}{\mbox{\boldmath $E$}}
\newcommand{\HH}{\mbox{\boldmath $H$}}
\newcommand{\V}{\mbox{\boldmath $V$}}
\newcommand{\G}{\mbox{\boldmath $G$}}
\newcommand{\W}{\mbox{\boldmath $W$}}
\newcommand{\St}{\mbox{\boldmath $S$}}
\newcommand{\Z}{\mbox{\boldmath $Z$}}

\newcommand{\Ypred}{\hat{\Y}}

\newcommand{\Gset}{\mbox{\boldmath $\mathcal{G}$}}
\newcommand{\video}{\mbox{\boldmath $v$}}
\newcommand{\ingredient}{\mbox{\boldmath $g$}}

\newcommand{\e}{\mbox{\boldmath $e$}}

\renewcommand{\L}{\mathcal{L}}
\newcommand{\Hset}{\mbox{\boldmath $\mathcal{H}$}}

\def\Bdma#1{\mbox{\boldmath{$#1$}}}

\renewcommand{\ss}{{v\rightarrow v}}
\newcommand{\ins}{{mat\rightarrow v}}

\newcommand{\biLSTM}{\mathrm{biLSTM}}

\newcommand{\VI}{{\sf VI} }
\newcommand{\VIV}{{\sf VIV} }
\newcommand{\VIVT}{{\sf VIVT} }

\definecolor{deepgreen}{rgb}{ .0, .44, .0}

\def\RepA#1#2{\textcolor{red}{$^{#1}$#2}}

\maketitle

\section{Introduction}
\label{sec:introduction}
With a rapid increase of cooking videos uploaded on the web, multimedia food computing has become an important topic \cite{salvador2017cvpr,salvador2019cvpr,pan2020acmmm,papadopoulos2022cvpr,wang2020eccv,zhang2021tmm}. Among various developing technologies, recipe generation from unsegmented cooking videos \cite{shi2019acl} is challenging. It requires artificial agents to (1) extract key events that are essential to dish completion and (2) generate sentences for the extracted events.
This task is important for both scene understanding and real-world applications. In terms of scene understanding, (1) and (2) are recognized as temporal event localization \cite{zhou2018aaai,escprcia2016eccv} and video captioning \cite{lei2020acl,shi2020acmmm,nishimura2021acmmm}, respectively, and learning both simultaneously is an ambitious challenge in computer vision (CV) and natural language processing (NLP).
Meanwhile, from a perspective of real-world applications, this technology can support people learning new skills by providing key events and their explanation sentences as a multimedia summarization of cooking videos. These academic and industrial motivations impel us to tackle recipe generation from unsegmented cooking videos.

Our task is similar to dense video captioning (DVC) \cite{krishna2017iccv,wang2021iccv,deng2021cvpr}, which aims at detecting events \textit{densely} from videos and generating sentences for them. Although the input/output pairs of our task (video/(events, sentences)) and that of the DVC are in the same format, our task is different from the perspective of the recipe's story awareness.
DVC allows agents to exclusively detect false-positive events in its evaluation metrics.
Our task, on the other hand, requires them to extract the accurate number of events in the correct order and generate sentences for them.
Fujita \etal \cite{fujita2020eccv} reported that DVC models produced more than 200 redundant events per video on average on the ActivityNet captions dataset \cite{krishna2017iccv} while the number of manually-annotated events is only 3-4.
Such redundant events and sentences make it difficult for users to grasp an overview of the video content.

\begin{figure}[t]
  \centering
  \includegraphics[width=0.8\linewidth]{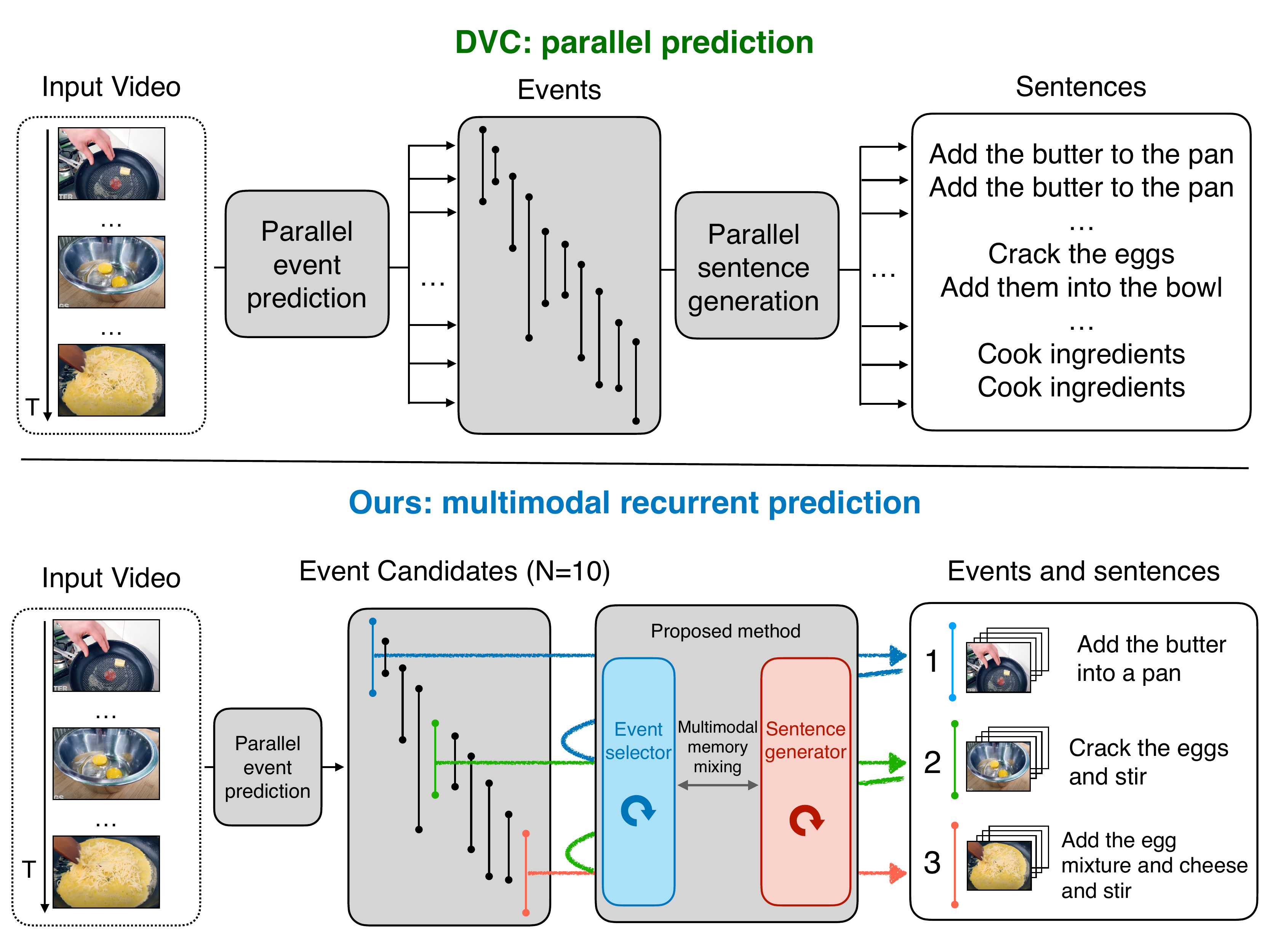}
  \caption{A conceptual comparison of our approach and existing DVC studies. While the existing DVC models adopted parallel prediction, our approach employ multimodal recurrent prediction, which estimates events and sentences by memorizing and fusing the previously prediction results.}
  \label{fig:task_overview}
\end{figure}

The reason for the redundant outputs is that existing DVC approaches employed parallel prediction, wherein events and sentences are estimated independently (\figref{fig:task_overview} top).
These approaches consist of two modules: event prediction and sentence generation modules. Both modules do not take into account the surrounding events and sentences when predicting the current ones, leading to duplicated outputs.
The choice of parallel prediction is reasonable for the DVC objective of thoroughly detecting events and generating sentences for multimedia retrieval.
However, it is not suitable for our recipe generation task because we place high importance on providing an overview of videos that can be easily grasped by humans.

Although the events predicted by DVC models are redundant, we observed that (1) several events are adoptable as a story of a recipe, but (2) the generated sentences for such events are not grounded well to the visual contents (i.e., ingredients and actions in the sentences are incorrect).
We confirm this by analyzing the outputs of the state-of-the-art DVC model.
We refer to the DVC output events as event candidates.
Through an approach that we refer to as oracle selection, we select events that have the maximum temporal IoU (tIoU) to ground-truth events (i.e., manually-annotated events) and compute the DVC scores \cite{fujita2020eccv,krishna2017iccv}.
The results support our observations.
Although oracle selection ensures that events that are adoptable as the story of a recipe are selected, the obtained sentences are not grounded in the visual content.

Based on this analysis, we set our goal to obtain correct recipes by selecting oracle events from the event candidates and re-generating sentences for them.
To realize this, the crucial idea is multimodal recurrent prediction, which estimates the next step by understanding previously predicted events and generated sentences.
The bottom of \figref{fig:task_overview} shows a concept of our idea.
To predict the events and sentences in step 3, the models should understand both the visual and textual content of step 2.
To achieve this, we propose a transformer\cite{vaswani2017neurips}-based recurrent learning approach of an event selector and sentence generator.
As with \cite{lei2020acl}, both of the modules have memory representations, which remember previously selected events and generated sentences to predict the next step effectively.
In addition, the proposed multimodal memory mixing method enables the modules to share the history of the previous prediction results, contributing the better performance.
The overall model is designed to be trainable in an end-to-end manner because the modules of the model are connected without breaking a differentiable chain. We refer to this model as the \textbf{base} model.

We also propose an \textbf{extended} model in the settings where the inputs are videos with ingredients as in \cite{nishimura2021acmmm,wu2022icmr}.
The motivation behind this extension is to enhance the model's ability to generate more accurate recipes by incorporating the actual ingredients depicted in the videos.
Relying solely on video inputs is problematic since even humans struggle to verbalize precise ingredients without additional contextual information.
For example, the kinds of seasoning (e.g., salt and sugar) or meat parts (e.g., chuck and rib) are the exact cases.
To develop a reliable recipe generation system, the models need to accept supplementary ingredient inputs.

This task requires agents to describe detailed manipulations of ingredients from cooking videos.
To achieve this goal, our extended model has two additional modules: (1) a dot-product visual simulator and (2) a textual attention module.
The dot-product visual simulator is introduced by the insights from our previous work \cite{nishimura2021acmmm}, where we discovered that the manipulations change the state of the ingredients, yielding state-aware visual representations. In \figref{fig:task_overview}, ``eggs'' are transformed into ``cracked'' and then ``stirred'' form to cook scrambled eggs.
To address this, we introduced the visual simulator, which reasons about the state transition of ingredients, into the model.
The motivation for the extension is the difference in the inputs.
Although the inputs in our previous work are pre-segmented ground-truth events with ingredients, the inputs in this work are event candidates.
The dot-product visual simulator accepts them and reasons about the state transition of ingredients.
Furthermore, we also introduce a textual attention module that verbalizes grounded ingredients and actions in the recipes. These modules are effective for grounded recipe generation from cooking videos.

In our experiments, we use the YouCook2 \cite{zhou2018aaai} dataset, which contains 2,000 videos with event/sentence annotation pairs. The proposed method outperforms the state-of-the-art DVC approaches based on commonly used DVC metrics. In addition, the extended models boost the model's performance. We also show that the proposed models can select the correct number of events that effectively reflects the ground-truth events.
In addition, our qualitative evaluation reveals that the proposed approaches can select events in the correct order and generate recipes grounded in the video contents. We discuss the detailed experiment settings for optimal recipe generation.

In summary, our contributions are summarized as follows:
\begin{itemize}
    \item This study focuses on enhancing the story awareness of recipes, where agents need to accurately extract the number of events in the correct order and generate corresponding sentences. Based on the analysis of the DVC models, we set our goal to obtain accurate recipes by selecting oracle events from the DVC output events and re-generating sentences for them.
    \item To achieve this goal, we propose transformer-based multimodal recurrent learning of an event selector and sentence generator. Building upon the predicted events from DVC models, our method effectively recurrently estimates the next step by memorizing and mixing previously predicted events and sentences.
    \item Furthermore, we extend our model to handle inputs comprising videos and ingredients. The motivation behind this extension is to enhance the model's ability to generate more precise recipes by incorporating actual ingredients observed in cooking videos. We introduce two additional modules to the base model, allowing it to verbalize detailed manipulations of ingredients in the cooking videos.
    \item Our experiments demonstrate that the proposed methods obtain the story awareness of a recipe. They outperform the DVC approaches especially on the story-oriented metrics, select the accurate number of events in the correct order, and generate accurate sentences for them. In addition, we investigate the detailed experiment settings for optimal recipe generation.
\end{itemize}

\section{Related Work}

\begin{table}[t]
\centering
\caption{A comparison of our methods with previous studies on recipe generation and video captioning. T, I, IS, S, P, and U represents text, images, image sequences, short event, pre-segmented sequential events, and unsegmented videos, respectively.}
\scalebox{0.8}{
\begin{tabular}{c|c|c|c}
\hline
Methods   & Input & Modalities            & Domain  \\ \hline
Kiddon \etal \cite{kiddon2015emnlp} & T & Title + ingredients & Cooking \\
Salvador \etal \cite{salvador2017cvpr} & I & Image & Cooking \\
Chandu \etal \cite{chandu2019acl} & IS & Image & Cooking \\ \hline
Shi \etal \cite{shi2020acmmm}       & S     & Video + transcription & Cooking \\
Wu \etal \cite{wu2022icmr}        & P     & Video + ingredients   & Cooking \\
Nishimura \etal \cite{nishimura2021acmmm} & P     & Video + ingredients   & Cooking \\
Lei \etal \cite{lei2020acl}       & P     & Video                 & General \\ \hline
Shi \etal \cite{shi2019acl}       & U     & Video + transcription & Cooking \\
Zhou \etal \cite{zhou2018cvpr}      & U     & Video                 & General \\
Wang \etal \cite{wang2021iccv}      & U     & Video                 & General \\
Ours      & U     & Video (+ ingredients) & Cooking \\ \hline
\end{tabular}
}
\label{tab:method_comparison}
\end{table}

In this section, we describe the novelty of the proposed method in line with other works on recipe generation from visual observations (\secref{subsec:recipe_generation}) and general video captioning (\secref{subsec:video_captioning}).
\tabref{tab:method_comparison} shows a comparison of our approaches with the previous studies.

\subsection{Recipe generation from visual observations}
\label{subsec:recipe_generation}
Recipes are a popular target in the NLP community because they require artificial agents to generate coherent sentences \cite{kiddon2016emnlp,bosselut2018naacl}. The inputs of these works are the title and ingredients of recipes.
Recently, many multimodal cooking datasets have appeared \cite{salvador2017cvpr,harashima2017sigir} and recipe generation from images has also been proposed \cite{salvador2019cvpr,wang2020eccv}.
Salvador \etal \cite{salvador2019cvpr} tackled this problem using a transformer-based model \cite{vaswani2017neurips} to generate high-quality recipes from a single image of a completed dish. Other researchers focus on generating a recipe from an image sequence depicting the intermediate food states \cite{chandu2019acl,nishimura2019inlg,nishimura2020ieee}.

The essential limitation of generating recipes from images is the lack of detailed, continuous scene information of human manipulations, which images do not contain.
Hence, this task is essentially an ill-posed problem; that is, it depends on the obtained language model to generate correct actions.
Meanwhile, videos contain this information and have been attracting the attention of many researchers in recent years \cite{nishimura2021acmmm,wu2022icmr,shi2019acl,shi2020acmmm}.
As described in \secref{sec:introduction}, this task requires agents to (1) extract key events from videos and (2) generate grounded sentences for events. Most researchers have focused on (2) generating grounded recipes from pre-segmented key events \cite{nishimura2021acmmm,shi2020acmmm,wu2022icmr}.
Our work differs from the studies categorized in (2) because our task requires the models to predict both (1) and (2) jointly.

However, only a few researchers attempt to learn both (1) and (2) to generate recipes from unsegmented videos.
Shi \etal \cite{shi2019acl} are pioneer researchers that have tackled this problem by utilizing the narration of videos effectively.
Although this approach is effective for narrated videos, transcription is not always available for all cooking videos.
Speaking during cooking or adding narration to videos requires significant effort. In addition, narrations may lead the model to attend to textual features, ignoring the visual features of videos. Our task extends their work to generate recipes only from videos. This enables the model to treat even non-narrated videos and be more useful than the previous setting.

\subsection{Video captioning}
\label{subsec:video_captioning}
Video captioning is an attractive field for both CV and NLP communities.
The task settings of video captioning vary according to the nature of the input video (e.g., one short event, pre-segmented sequential events, or long unsegmented video). 
We aim to extract key events and generate sentences simultaneously from long unsegmented videos.
This task is similar to DVC because the input/output pairs are in the same format as in our task. We first describe traditional DVC approaches and then enumerate the difference between DVC and our task.

DVC aims at densely detecting events from the video and generating sentences for them. Recently, transformer-based approaches are well investigated for this task. Zhou \etal \cite{zhou2019cvpr} proposed a Masked Transformer, which detects events and generates sentences via a differentiable mask in an end-to-end manner. Their approach, however, outputs more than 200 redundant events per video on average \cite{fujita2020eccv}. To handle this issue, Wang \etal \cite{wang2021iccv} proposed PDVC, which detects events in parallel, re-ranks the top $K$ ($K$ is also the prediction target), and generates sentences for the re-ranked events.
Deng \cite{deng2021cvpr} proposed a top-down DVC approach, whereby whole sentences generated for the video are aligned into timestamps, based on which the sentences are refined.

The key difference between our task and DVC lies in the story awareness of recipes. Our task requires agents to extract an appropriate number of events in the correct order.
DVC allows a model to detect events densely; however, redundant event/sentence pairs impede readability, and users are unable to grasp an overview of the video contents.
To achieve story awareness, we propose a transformer-based joint learning approach of an event selector and sentence generator based on our observation that oracle events can be selectable from the output events of a DVC model.

\section{Oracle-based Analysis of the Existing DVC Model}
\label{sec:oracle_analysis}

\begin{table}[t]
\centering
\caption{Word-overlap metrics of the oracle selection on the YouCook2 dataset. $N$ represents the number of candidate events, a hyper-parameter of PDVC. The bold scores are the best among the comparative settings.}
\scalebox{0.8}{
\begin{tabular}{c|ccc|ccc}
\hline
& \multicolumn{3}{|c|}{dvc\_eval} & \multicolumn{3}{c}{SODA} \\ \hline
& BLEU4 & METEOR & CIDEr-D & METEOR & CIDEr-D & tIoU \\ \hline
PDVC & 0.89 & 4.52 & 21.50 & 3.98 & 25.30 & 27.80 \\ \hline
Oracle & & & & & & \\ \hline
N=25 & 0.58 & 6.09 & 27.12 & 7.62 & 26.32 & 56.55 \\
N=50 & 0.84 & 6.92 & 31.63 & 8.83 & 29.93 & 64.58 \\
N=100 & 0.97 & 7.68 & 36.26 & 9.64 & 35.08 & 71.16 \\
N=200 & \textbf{1.10} & \textbf{8.15} & \textbf{38.60} & \textbf{10.43} & \textbf{36.89} & \textbf{76.71} \\ \hline
\end{tabular}
}
\label{tab:oracle_dvc}
\end{table}

\begin{figure}[t]
  \centering
  \includegraphics[width=0.8\linewidth]{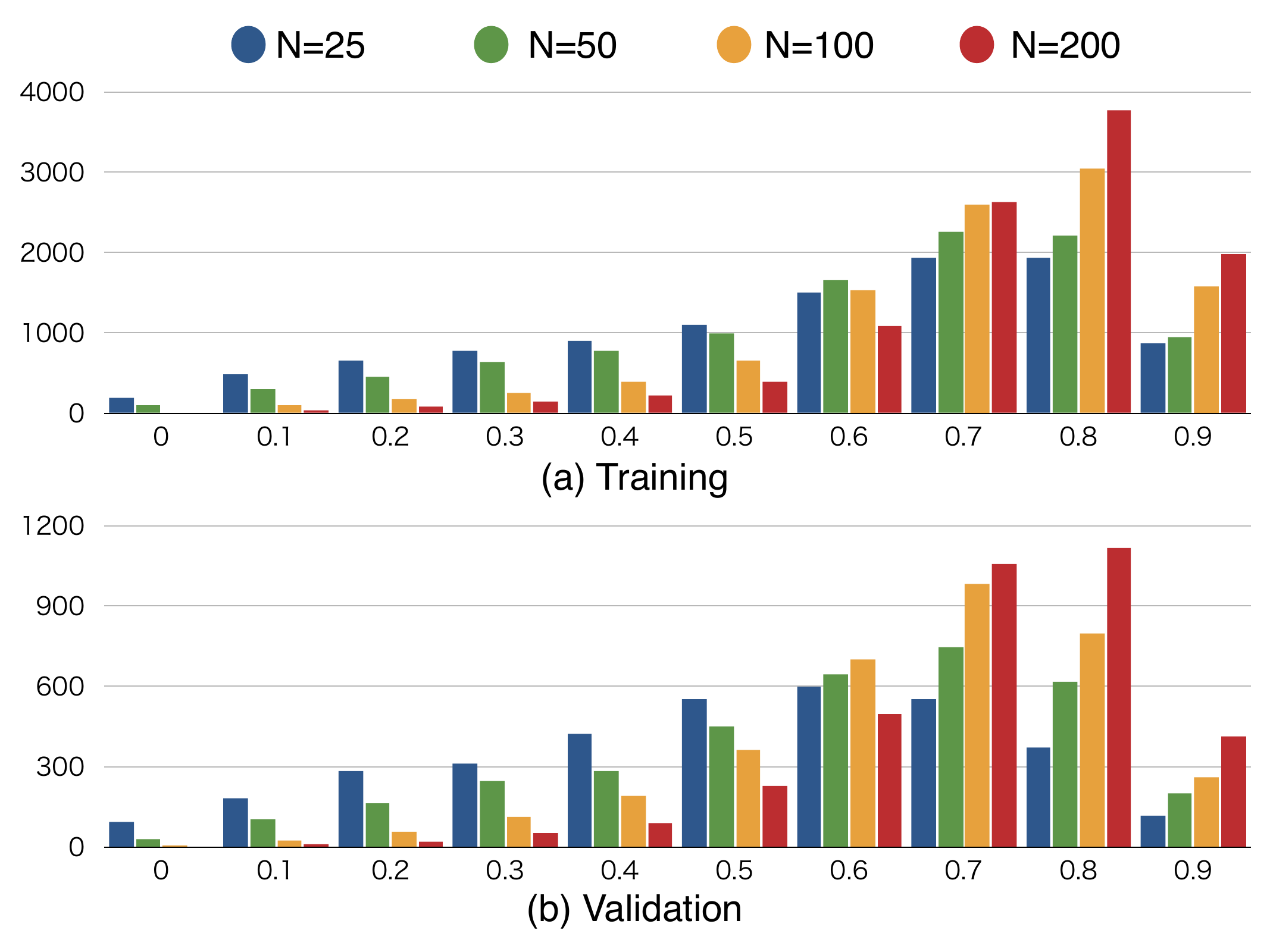}
  \caption{tIoU distribution of oracle events on the training and validation sets of YouCook2.}
  \label{fig:oracle_analysis}
\end{figure}

As Fujita \etal \cite{fujita2020eccv} reported, the outputs of DVC models are redundant. However, we observed that although (1) several events are adoptable as a story of a recipe, (2) the generated sentences for such events are not grounded well in the visual contents.
To verify this, we analyze the outputs of the state-of-the-art DVC model. Specifically, we select events with the maximum tIoU scores to ground-truth events from event candidates and compute the DVC scores, an approach referred to as oracle selection hereinafter.
The analyzed results demonstrate that the oracle selection boosts the performance, especially on story-oriented DVC metrics.

We use the YouCook2 dataset \cite{zhou2018aaai}, one of the largest cooking video-and-language datasets. For the DVC model, we employ PDVC \cite{wang2021iccv}, the state-of-the-art DVC model. It detects $N$ events densely and then re-ranks the top $K$ of them for its outputs.
Note that $N$ is a hyper-parameter\footnote{$N$ is set to be a sufficiently large number. For YouCook2, the authors of \cite{wang2021iccv} set $N$ to be $100$ as a default parameter.} and $K$ is the prediction target.
We adopt this model because it achieves the best performance on DVC tasks and it is easy to control $N$ before training the model. We use the $N$ detected events for the analysis, not the re-ranked $K$ events.

\textbf{Evaluation metrics.}
We use two commonly-used DVC metrics: dvc\_eval \cite{krishna2017iccv} and SODA \cite{fujita2020eccv}.

\begin{itemize}
    \item \textbf{dvc\_eval} firstly computes tIoU of all the combinations between the prediction and ground-truth events and then computes word-overlap metrics (e.g., METEOR or CIDEr-D), if the tIoU scores are over the threshold $\theta$. $\theta$ ranges from $0.3$ to $0.9$ by $0.2$. Their average is the output score of these metrics.
    \item \textbf{SODA} stands for story-oriented dense video captioning evaluation, whereby the story awareness of the output events is evaluated. Specifically, it uses dynamic programming to explore an alignment of events between prediction and ground truth for obtaining the maximum story scores. The story scores are computed as a product of tIoU and word overlap metrics. Because SODA evaluates the predicted events by penalizing redundant outputs, it is suitable for computing the story awareness of recipes.
\end{itemize}

SODA evaluates whether the output event/sentence pairs are appropriate as a story; thus, it is rated above dvc\_eval in this study.
As word-overlap metrics, we use BLEU \cite{papineni2002acl}, METEOR \cite{banerjee2005acl}, and CIDEr-D \cite{vedantam2015cvpr}, which are commonly-used metrics for text generation tasks. We also introduce SODA tIoU, which computes story scores as tIoU scores, rather than a product of tIoU and word overlap metrics. These metrics can evaluate whether the selected events are appropriate as components of the generated recipes. In this analysis, we range $N$ from $25, 50, 100,$ and $200$.

\subsection{Quantitative evaluation}
\tabref{tab:oracle_dvc} shows the results of the oracle selection on dvc\_eval and SODA metrics, indicating that the oracle selection outperforms the PDVC. Specifically, the SODA scores of the oracle selection are quite better than that of the PDVC, demonstrating that the generated recipes are more suitable as a story than the ones generated by the PDVC.

\figref{fig:oracle_analysis} shows the distribution of tIoU scores of the oracle events on training and validation sets.
The number of candidate events $N$ was directly proportional to the average tIoU. This is because the more $N$ was, the more suitable oracle events appear in the candidate. Both the training and validation sets confirm this tendency.

\subsection{Qualitative evaluation}

\begin{figure}[t]
  \centering
  \includegraphics[width=0.8\linewidth]{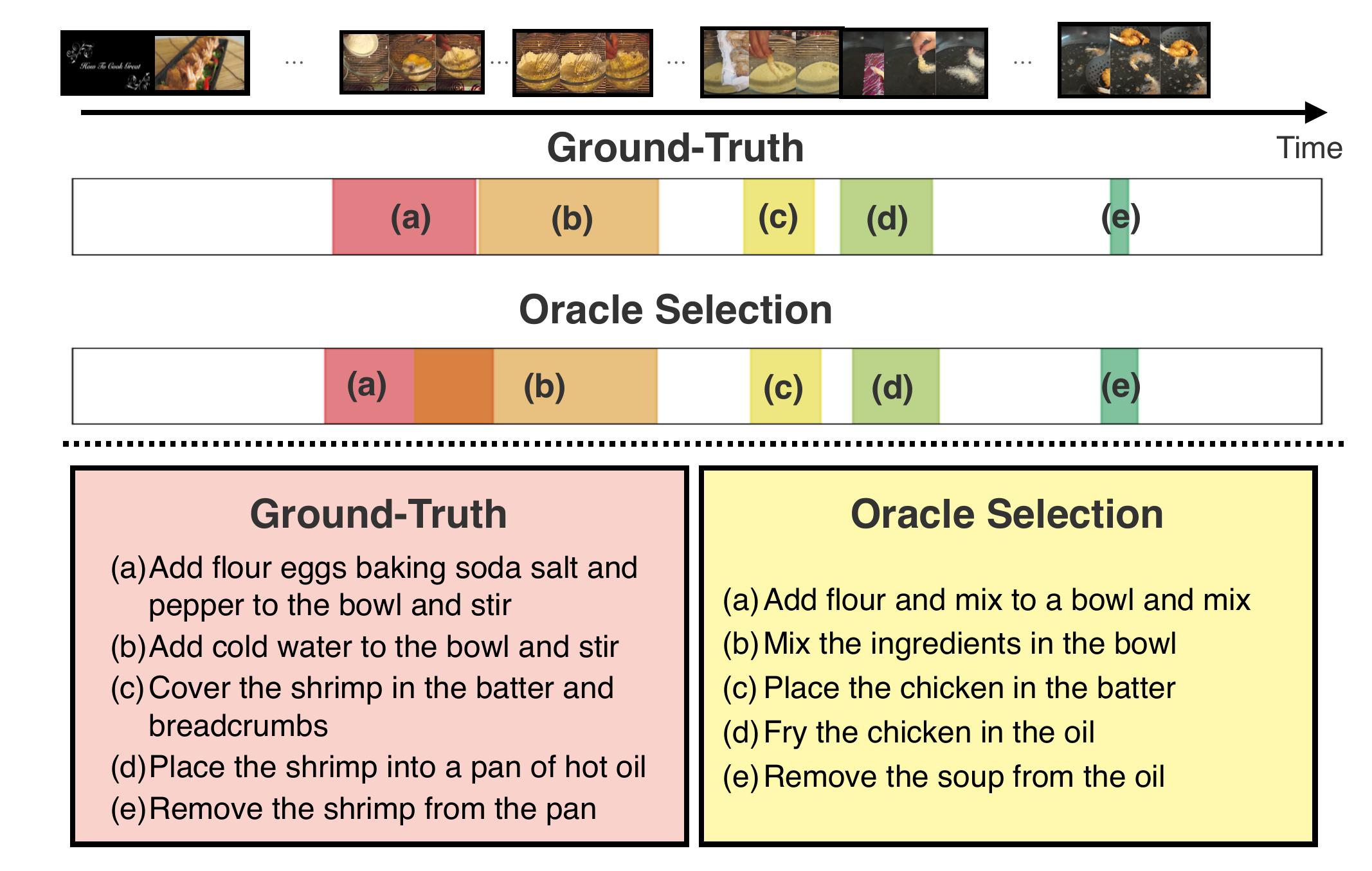}
  \caption{Comparison of the recipes generated by the oracle selection and ground truth. $N=100$, which is a default hyper-parameter of PDVC, is used in this example.}
  \label{fig:oracle_recipes}
\end{figure}

\figref{fig:oracle_recipes} shows a comparison of the recipes generated by the oracle selection and ground truth. The selected event timestamps are close to the ground-truth events, indicating that the appropriate selection can construct the correct recipes. However, the sentences are different from the ground truth (e.g., ``baking soda'', ``salt'', and ``pepper'' are missing in step (1)).

The main reason for this error is parallel prediction, where the events and sentences are predicted independently.
This causes the model to miss or hallucinate ingredients and motivates us to propose a model that re-generates sentences for the predicted events, rather than using the corresponding generated sentences without modification.

\section{Proposed method}
\label{sec:proposed_method}

\begin{figure}[t]
  \centering
  \includegraphics[width=0.8\linewidth]{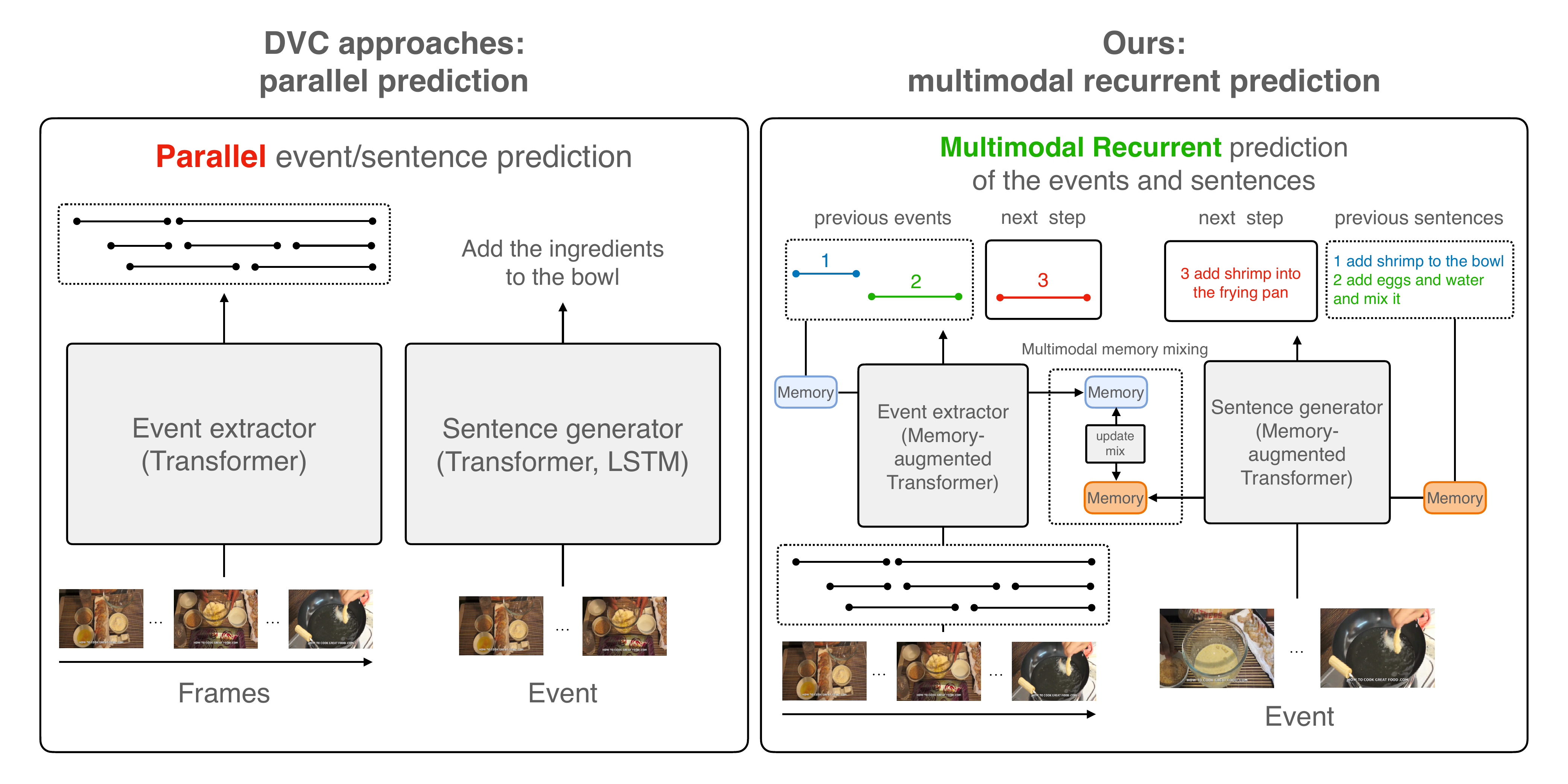}
  \caption{An introductory overview of our approach. Unlike the previous DVC approaches, we propose a multimodal recurrent learning approach to train the event selector and sentence generator. Both modules represent the previously predicted events and sentences as memory vectors and predict the next step. These memory vectors are updated and mixed to effectively share the previous prediction belonging to different modalities.}
  \label{fig:introductionary_figure}
\end{figure}

\begin{figure}[t]
  \centering
  \includegraphics[width=0.8\linewidth]{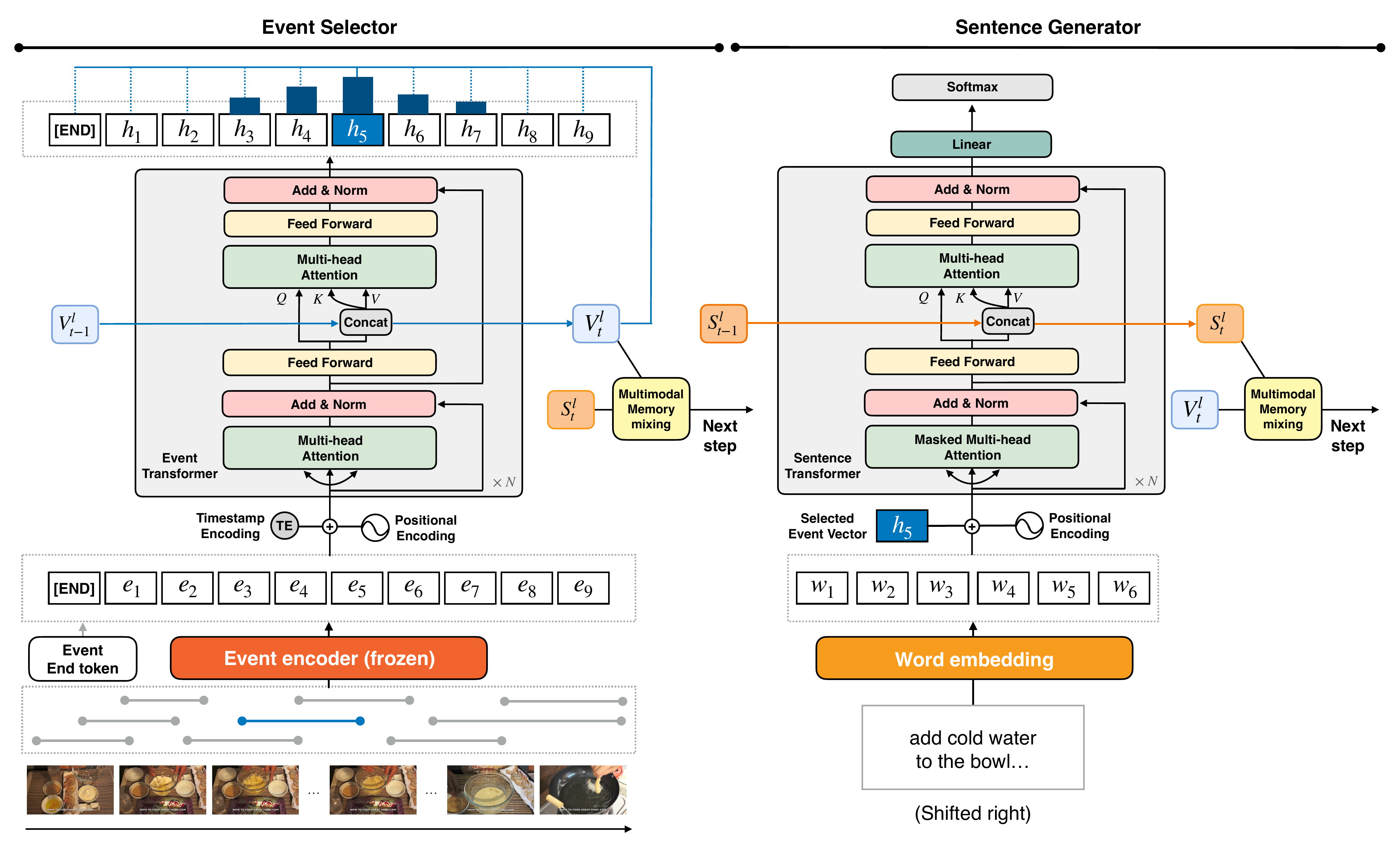}
  \caption{Multimodal recurrent learning approach of the event selector and sentence generator for recipe generation from unsegmented cooking videos. The event selector tries to choose oracle events from event candidates repeatedly (\secref{subsec:event_selector}) and the sentence generator outputs sentences for the selected events (\secref{subsec:sent_generator}).
  The memories are updated and mixed to effectively remember the history of the events/sentences for predicting the next step (\secref{subsec:memory_update}).}
  \label{fig:proposed_method}
\end{figure}

Based on the oracle-based analysis, we set our goal to obtain correct recipes by selecting oracle events from the output events of the DVC model and re-generating sentences for them.
To achieve this goal, multimodal recurrent prediction is essential, which memorizes and mixes the previously predicted events and generated sentences to estimate the next step (\figref{fig:introductionary_figure}).

We realize this idea by proposing transformer-based models, consisting of event selection and sentence generation modules (\figref{fig:proposed_method}), which we refer to them as an event selector and sentence generator.
The event selector tries to choose oracle events from event candidates repeatedly (\secref{subsec:event_selector}) and the sentence generator outputs sentences for the selected events (\secref{subsec:sent_generator}).
Both encoders are based on memory-augmented recurrent transformers (MART) \cite{lei2020acl}, which contain memory vectors to learn the recurrence by remembering the previous prediction to estimate the next step.
In addition, the proposed multimodal memory mixing approach enables the model to share the previous prediction belonging to different modalities by fusing their memory vectors (\secref{subsec:memory_update}).

It is worth noting the advancements over MART.
This study presents two improvements:
\begin{itemize}
    \item \textbf{Task extension.} We extend the scope of the model to verify both event selection and sentence generation tasks, whereas the original paper only validated its effectiveness in generating sentences from pre-segmented key events.
    \item \textbf{Multimodal memory mixing.} MART contains one memory vector to tackle only the sentence generation, but two memory vectors are necessary for our task, which predicts events and sentences jointly.
    How to treat multiple memory vectors is not investigated in MART, thus we first propose a memory mixing method that integrates them from both modules achieve accurate recipe generation (\secref{subsec:memory_update}).
\end{itemize}

Mathematically, we formulate our task using notations. Let $\X=(\x_1, \x_2, \ldots, \x_n, \ldots, \x_N) \in \mathbb{R}^{2 \times N}$ be event candidates, where $\x_n$ consists of start and end timestamps. Note that $\X$ is sorted on the start time of the events in chronological order.
Given $\X$, the model generates pairs $(\C,\Y)=((c_1,\y_1), \ldots, (c_t, \y_t), \ldots, (c_T, \y_T))$, where $c_t$, $\y_t$, and $T$ represent an index of the oracle event candidates, corresponding generated sentences, and the number of the selected events, respectively.
The memories $\V_t^{l}$ in the event selector and $\St_t^{l}$ in the sentence generator are jointly updated at each $t$ step, where $l$ represents the layer number of transformers.

\subsection{Event selector}
\label{subsec:event_selector}
The event selector can be divided into two main components: event encoder and event transformer. The event encoder converts given candidate events $\X$ into event-level representations $\E=(\e_1, \e_2, \ldots, \e_n, \ldots, \e_T)$, based on which the event transformer outputs holistic representations of events $\HH=(\h_1, \h_2, \ldots, \h_n, \ldots, \h_T)$.
They are used for computing event probabilities, which represent the likelihood of oracle events. We apply Gumbel softmax resampling \cite{jang2017iclr} to select events and forward them to the sentence generator without breaking a differentiable chain during the training phase. In the inference phase, events are deterministically selected by applying argmax to event probabilities.

\textbf{Event encoder.} The event encoder converts events $\x_n$ into representations $\e_n$. First, we extract event clips from the video, according to the start and end time of $\x_n$, and input them into pre-trained visual encoders. One of the straightforward ways to encode events is to average frame-level image representations extracted by ResNet \cite{he2016cvpr}. However, we find that this approach is unable to capture fine-grained event semantics for overlapped events. For example, assume that the ground-truth event is a scene of cutting potatoes, and the event candidates have two scenes, where one is a scene of cutting potatoes and tomatoes, and another is of cutting only potatoes. We expect the model to select the latter scene because the former contains extra information about cutting tomatoes. Representing events based on the average of frame-level features cannot capture the semantic difference of these events effectively.
Thus, it is necessary to employ an event encoder that focuses on extracting fine-grained event-level semantics.

To achieve this, we focus on the multiple instance learning-noise contrastive estimation (MIL-NCE) model \cite{miech2020cvpr} pre-trained on Howto100M \cite{miech2019iccv}. The model is trained on more than 100M pairs of an event with narration, and it can capture event-level semantics of overlapped events.
Using this encoder, we obtain event-level representations as $\E=(\e_1, \e_2, \ldots, \e_n, \ldots, \e_T)$. Then, the positional encoding (PE) \cite{vaswani2017neurips} and relative encoding of events \cite{wang2021tcsvt} are added to $\e_n$.
Note that the event encoders are pre-trained and their weights are frozen during the training phase because of the computational costs.

\textbf{Event transformer.} Given $\E$, the event transformer outputs holistic representations of events $\HH=(\h_1, \h_2, \ldots, \h_n, \ldots, \h_T)$. This module is based on the memory-augmented recurrent transformer \cite{lei2020acl}, where each $l$ layer has the memories $\V_t^l$ to remember the history of the selected events. The output vectors $\HH=(\h_1, \h_2, \ldots, \h_n, \ldots, \h_T)$ and memories $\V_t^l$ are used to compute $n$-th event probability $p(c_t=n|\V_{t}^{l},\X)$ as follows:
\begin{eqnarray}
    \V_{t} &=& \max(\V_{t}^{1}, \ldots, \V_{t}^{l}, \ldots, \V_{t}^L), \\
    p(c_t=n|\V_{t}^{l},\HH) &=& \frac{\exp{\{(\h_n^t)^\mathrm{T}\V_t\}}}{\sum_{i}\exp{\{(\h_i^t)^\mathrm{T}\V_t\}}},
\label{eq:event_selection}
\end{eqnarray}
where $\max(\cdot)$ represents an element-wise max-pooling of vectors. During training, the event selector selects events through a Gumbel softmax resampling \cite{jang2017iclr}, which enables us to train the model in an end-to-end manner without breaking a differentiable chain. This indicates that the loss computed on the sentence generator is backpropagated to the event selector. During inference, the model selects the event index based on the argmax of $p(c_t=n|\V_{t}^{l},\HH)$.

\subsection{Sentence generator}
\label{subsec:sent_generator}
Based on the selected event representations, the sentence generator outputs sentences grounded for them. Let the selected event index be $\hat{c}_t$, and the selected event representation can be written as $\h_{\hat{c}_t}$. This vector is added to the word vectors $\W=(\w_1, \w_2, \ldots, \w_k, \ldots, \w_K)$ from the word embedding layer, which is the concatenated neural networks of the pre-trained global vectors for word representation (GloVe) \cite{pennington2014emnlp}\footnote{We employ pre-trained 300D word embedding, which can be downloaded from \url{http://nlp.stanford.edu/data/glove.6B.zip}} and one-layer perceptron with ReLU activation. We also add PE to word vectors $\W$ and input them into the sentence transformer, another memory-augmented recurrent transformer. The model generates words repeatedly by applying softmax and argmax operations to the output vectors of the sentence transformer.

\subsection{Multimodal memory mixing}
\label{subsec:memory_update}
One of the important contributions of this study is to mix the memories for sharing the prediction results between the event selector and sentence generator.
The motivation of this approach is that mixing memories is intuitively effective for accurate recipe generation because the history of the selected events contributes to sentence generation and vice versa.
Specifically, the memory vectors $\V_t^l, \St_t^l$ are first separately updated by following the equations described in Section 3.2 in MART paper, and then mixed as follows:
\begin{eqnarray}
    \hat{\V}_t &=& f_1(\V_t) \odot \sigma(g_2(g_1(\St_t))), \\
    \hat{\St}_t &=& g_1(\St_t) \odot \sigma(f_2(f_1(\V_t))),
\end{eqnarray}
where $f_{*}(\cdot),g_{*}(\cdot)$ represents a single linear layer and $\odot,\sigma$ represents the Hadamard dot product and sigmoid function, respectively.
The obtained $\hat{\V}_t$ and $\hat{\St}_t$ are forwarded into the next $t+1$ step.
Note that we compare this with the original MART algorithm that updates memory vectors separately, and confirm its effectiveness in \secref{subsub:seperate_joint}.

\subsection{Loss functions}
\label{subsec:loss_function}
To train the model in an end-to-end manner, we sum up two types of losses: event selection loss and sentence generation loss. These losses are formulated as a negative log-likelihood of the event selection and sentence generation tasks as follows:
\begin{eqnarray}
    L_{base} &=& L_e + L_s, \\
    L_e &=& - \sum_{(\X,\C)}\sum_{t}\log{p(c_t|\X, \V_{<t}^l)}, \\
    L_s &=& - \sum_{(\X, \C, \Y)}\sum_{t}\log{p(\y_t|\h_{\hat{c}_t}, \St_{<t}^l)},
\end{eqnarray}
where $L_{base}$ represents the total loss of the event selection and sentence generation losses $L_e, L_s$.

\section{Extended model}

\begin{figure}[t]
  \centering
  \includegraphics[width=0.8\linewidth]{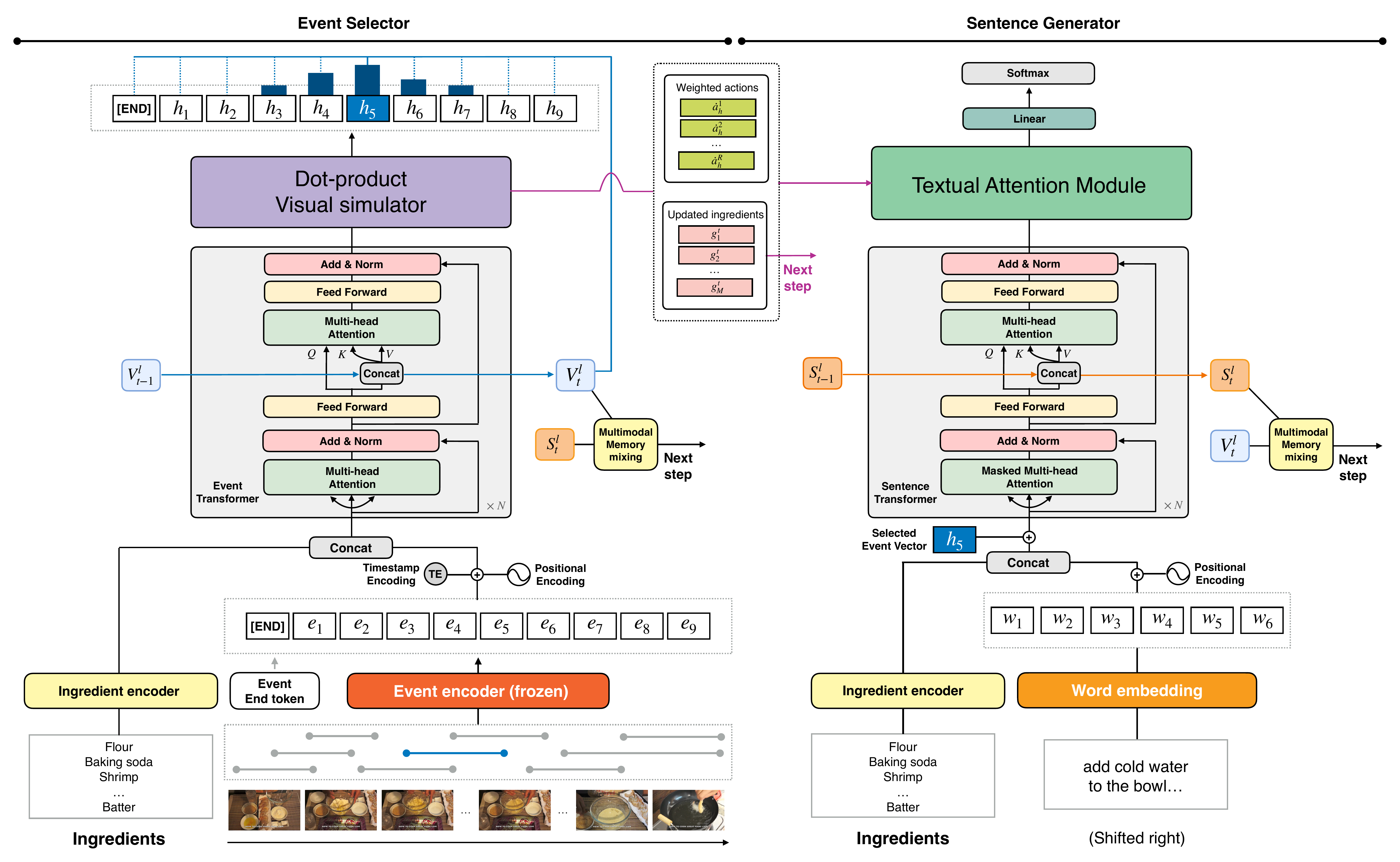}
  \caption{Overview of the extended model for recipe generation from unsegmented cooking videos. To generate more accurate recipes, it has additional two modules: (1) dot-product visual simulator and (2) textual attention. The dot-product visual simulator is introduced by the insights based on our previous work \cite{nishimura2021acmmm} to learn the state transition of ingredients. The textual attention module encourages the sentence generator to verbalize actions and ingredients more accurately.}
  \label{fig:extended_method}
\end{figure}

\begin{figure}[t]
  \centering
  \includegraphics[width=0.8\linewidth]{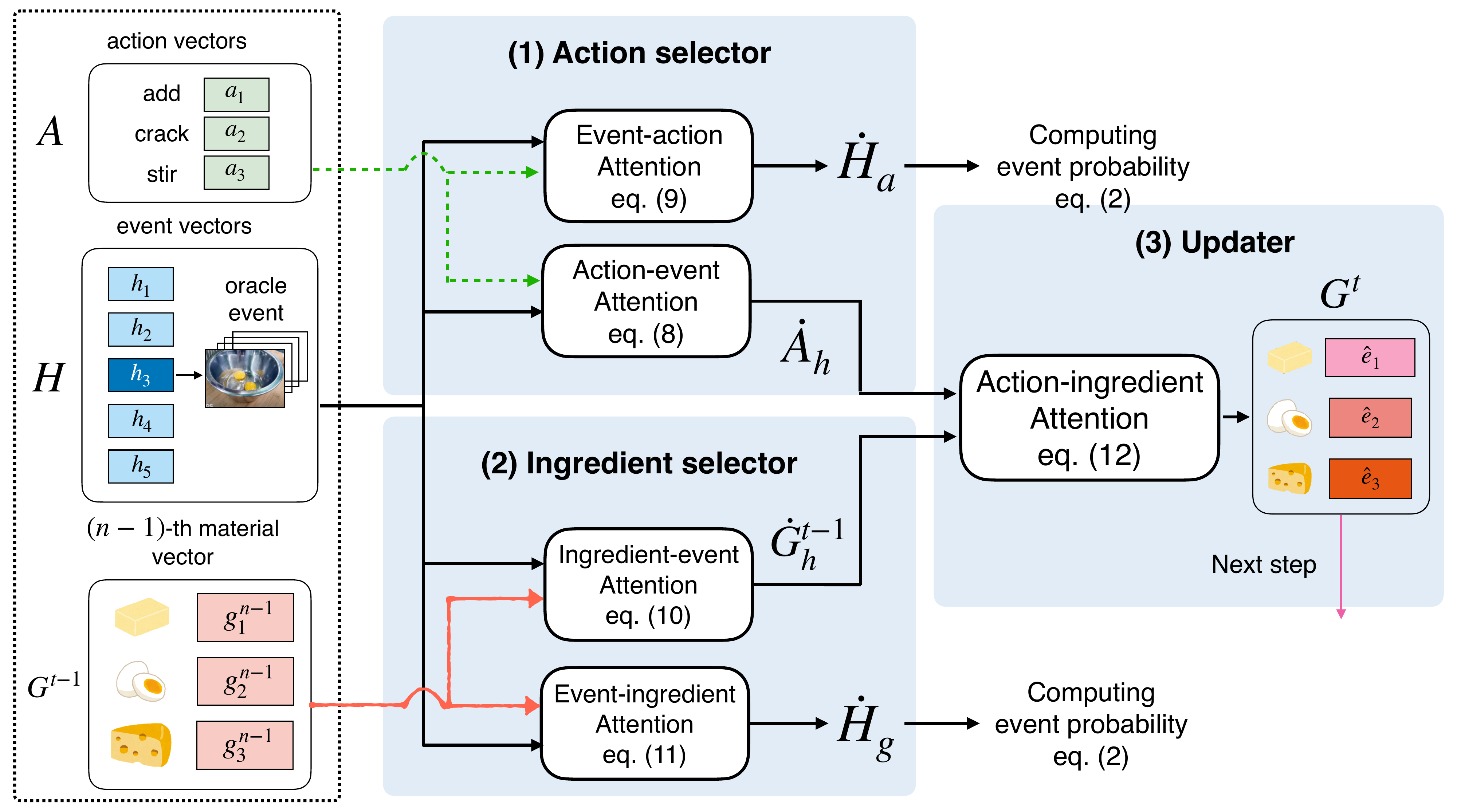}
  \caption{Overview of the extended dot-product visual simulator that reasons about the state transition of ingredients. It has three components: (1) action selector, (2) ingredient selector, and (3) updater. The dot-product attention is employed to treat the event candidates.}
  \label{fig:dot_product}
\end{figure}

In addition to the base model, we also propose an extended model in the settings where the inputs are videos with ingredients, as in \cite{nishimura2021acmmm,wu2022icmr}.
The motivation behind this extension is to enhance the model's ability to generate more accurate recipes by incorporating the actual ingredients depicted in the videos.
Relying solely on video inputs is problematic since even humans struggle to verbalize precise ingredient names without additional contextual information.
In \figref{fig:generated_recipe}, for instance, the correct ingredient is ``pork,'' yet the base model mistakenly generates ``meat'' and ``veal.''
To develop a reliable recipe generation system, the models need to accept supplementary ingredient inputs.

The challenge is to describe detailed manipulations of ingredients from cooking videos.
To achieve this goal, our extended model has two additional modules: (1) dot-product visual simulator and (2) textual attention module (\figref{fig:extended_method}).
The dot-product visual simulator is introduced based on the insights from our previous work \cite{nishimura2021acmmm}. In our previous work, we discovered that the manipulations change the state of the ingredients, yielding state-aware visual representations. For example, ``eggs'' are transformed into ``cracked,'' then ``stirred'' forms to cook scrambled eggs. To address this, we introduced the visual simulator into the model, which reasons about the state transition of ingredients.
Although pre-segmented ground-truth events with ingredients are regarded as input in our previous work, they are considered event candidates with ingredients in this study.
Thus we extend it to the dot-product visual simulator described in \secref{subsec:dot_product_vis_attn}. In addition, we also introduce a textual attention module that verbalizes grounded ingredients and actions in the recipes. These modules are effective for grounded recipe generation from cooking videos.

Another extension is to add an ingredient encoder to convert ingredients into representations $\G^0$. The ingredient encoders are concatenated neural networks of pre-trained GloVe \cite{pennington2014emnlp} word embedding and multi-layer perceptrons (MLPs) with ReLU activation function and are added to the event selector and sentence generator without sharing their parameters.
Multi-word ingredients (e.g., parmesan cheese) are represented by the average embedding vector of the words. They are concatenated with event/word representations and inputted to the event/sentence transformers as shown in \figref{fig:extended_method}.

\subsection{Dot-product visual simulator}
\label{subsec:dot_product_vis_attn}
We first revisit the original visual simulator and then describe how we extend it to the dot-product visual simulator.

\textbf{Visual simulator revisit.}
In the original study, the inputs are the pairs of ingredients and ground-truth events. Let $\G^0=(\g_1^0, \g_2^0, \ldots, \g_m^0, \ldots, \g_M^0)$ and $\hat{\HH}=(\hat{\h}_1, \hat{\h}_2, \ldots, \hat{\h}_{\hat{t}}, \ldots, \hat{\h}_{\hat{T}})$ be the ingredient and ground-truth event vectors encoded by ingredient encoder (e.g., GloVe) and event encoder (e.g., MIL-NCE), respectively.
Given $(\G^0, {\hat{\HH}})$, the visual simulator reasons about the state transition of the ingredients by updating them at each $\hat{t}$-th step. To this end, it consists of three components: (1) action selector, (2) ingredient selector, and (3) updater. The action selector and ingredient selector predict executed actions and used ingredients at $\hat{t}$-th step. Based on the selected actions/ingredients, the $(\hat{t}-1)$-th ingredient vectors $\G^{(\hat{t}-1)}$ are updated into $\hat{t}$-th new proposal ingredient vectors $\G^{(\hat{t})}$, which are forwarded into the next step. The visual simulator recurrently repeats the above process until processing the end element of the ground-truth events.

\textbf{Proposed extension.}
The proposed model is similar to the visual simulator because it also has the same three components to update the ingredient vectors. However, instead of the ground-truth event vectors $\hat{\HH}$, the event candidates $\HH$ are assumed to be the inputs in this study, that is, the model needs to predict not only the actions/ingredients but also the events that constitute the recipe story.

To achieve this, we extend the visual simulator into the dot-product visual simulator shown in \figref{fig:dot_product}. It computes the relationships between action-to- and ingredient-to-events as attention matrices and outputs four vectors: action-weighted and ingredient-weighted event vectors, event-weighted action, and ingredient vectors. The former two vectors are used to calculate the event probabilities and the latter two vectors are forwarded to the textual attention module.

\textbf{(1) Action selector} outputs event-weighted action and action-weighted event vectors by predicting the executed actions and related events.
Let the action vectors be $\A=(\ai_1, \ai_2, \ldots, \ai_r, \ldots, \ai_R)$, that is, the pre-defined action embedding, where $\ai_r$ represents $r$-th actions and $R$ is the number of the actions. In \figref{fig:dot_product}, the actions ``crack'' and ``stir'' are executed at the oracle event $\h_3$; thus $\ai_{\ast}$ that corresponding crack and stir indices should be selected with the relation to $\h_3$.
This computation can be formulated as the dot-product attention as follows:
\begin{equation}
    \hat{\A}_h = \mathrm{softmax}\{\frac{(\W_a^Q\A)^T(\W_h^K\HH)}{\sqrt{d}}\},\ \ \ \ \dot{\A}_h = \hat{\A}_h(\W_h^V\HH)^T,
\label{eq:adot}
\end{equation}
where $\W_a^Q, \W_h^K, \W_h^V$ represents a linear layer and $d$ represents the dimension size of $\A$ and $\HH$\footnote{The dimension size of $\A$ is set to be equal to $\HH$.}. Note that $\mathrm{Softmax(\cdot)}$ represents the row-wise softmax operation on a matrix.
We also acquire action-weighted event vectors $\dot{\HH}_a$ as follows:
\begin{equation}
    \dot{\HH}_a = \mathrm{softmax}\{\frac{(\W_h^Q\HH)^T(\W_a^K\A)}{\sqrt{d}}\}(\W_a^V\A)^T,
\label{eq:vdot}
\end{equation}
where $\W_h^Q, \W_a^K, \W_a^V$ represents a linear layer.

\textbf{(2) Ingredient selector} outputs the event-weighted ingredient and ingredient-weighted event vectors by predicting the ingredients used and related events.
For example, in \figref{fig:dot_product}, the raw ``eggs'' should be selected at $\h_3$.
This computation is achieved by replacing $\A$ with $\G^{t-1}$ in \eqref{eq:adot} and \eqref{eq:vdot} as follows:
\begin{align}
    \hat{\G}_h^{t-1} &= \mathrm{softmax}\{\frac{(\W_g^Q\G^{t-1})^T(\W_h^K\HH)}{\sqrt{d}}\},\ \ \ \ \dot{\G}_h^{t-1} = \hat{\G}_h^{t-1}(\W_h^V\HH)^T, \label{eq:gdot} \\
    \dot{\HH}_g &= \mathrm{softmax}\{\frac{(\W_h^Q\HH)^T(\W_g^K\G^{t-1})}{\sqrt{d}}\}(\W_g^V\G^{t-1})^T,
\end{align}
where $\W_g^Q, \W_g^K, \W_g^V$ represents a linear layer.

\textbf{(3) Updater} represents the state transition of the ingredients by updating the ingredient vectors $\G^{t-1}$. Based on the selected actions and ingredients $(\dot{\A}_h, \dot{\G}_H^{t-1})$, the updater computes the updated ingredient representations as follows:
\begin{equation}
    \G^t  = \G^{t-1} + \dot{\G}_h^{t-1} \odot \mathrm{repeat}(\mathrm{max}(\dot{\A}_h)),
\end{equation}
where $\mathrm{repeat}(\cdot)$ expands the max-pooled action vector by repeating it $M$ times ($M$ is the number of ingredients).

\textbf{Output representations.} The action- and ingredient-weighted event representations $(\dot{\HH}_a, \dot{\HH}_g)$ are used to compute the event probability by replacing $\HH$ with $\dot{\HH} = \HH + \dot{\HH}_a + \dot{\HH}_g$ in \eqref{eq:event_selection}.
The event-weighted action vectors $\dot{\A}_h$ and updated ingredients $\G^t$ are forwarded to the textual attention module. $\G^t$ is also set to be the ingredient vectors at $(t+1)$-th prediction.

\subsection{Textual attention}
The textual attention module encourages the sentence generator to output a recipe grounded with the events based on the actions and ingredients.
Let $\hat{\W}=(\w_1, \w_2, \ldots, \w_k, \ldots, \w_K)$ be the output word vector sequence from the sentence transformer.
Given $(\G^t, \dot{\A}_h)$ and $\hat{\W}$, the textual attention module computes two attention matrices: (1) word-ingredient attention and (2) word-action attention.
Then it outputs the context vectors $\hat{\U}$ as follows:
\begin{eqnarray}
    \Z_m &=& \mathrm{softmax}(\frac{\hat{\W}^T\G^t}{\sqrt{d}}),\ \ \ \ \U_m = \Z_m(\G^t)^T, \label{eq:word_gdot} \\
    \Z_a &=& \mathrm{softmax}(\frac{\hat{\W}^T\dot{\A}_h}{\sqrt{d}}),\ \ \ \ \U_a = \Z_a(\dot{\A}_h)^T, \label{eq:word_adot} \\
    \hat{\U} &=& \mathrm{concat}(\hat{\W}, (\U_m)^T, (\U_a)^T),
\end{eqnarray}
where $\mathrm{concat}(\cdot)$ indicates a concatenation function of vectors. $\hat{\U}$ is forwarded to the linear layer and softmax activation to obtain the word probability across the vocabulary.

\subsection{Loss functions}

\begin{figure}[t]
  \centering
  \includegraphics[width=0.8\linewidth]{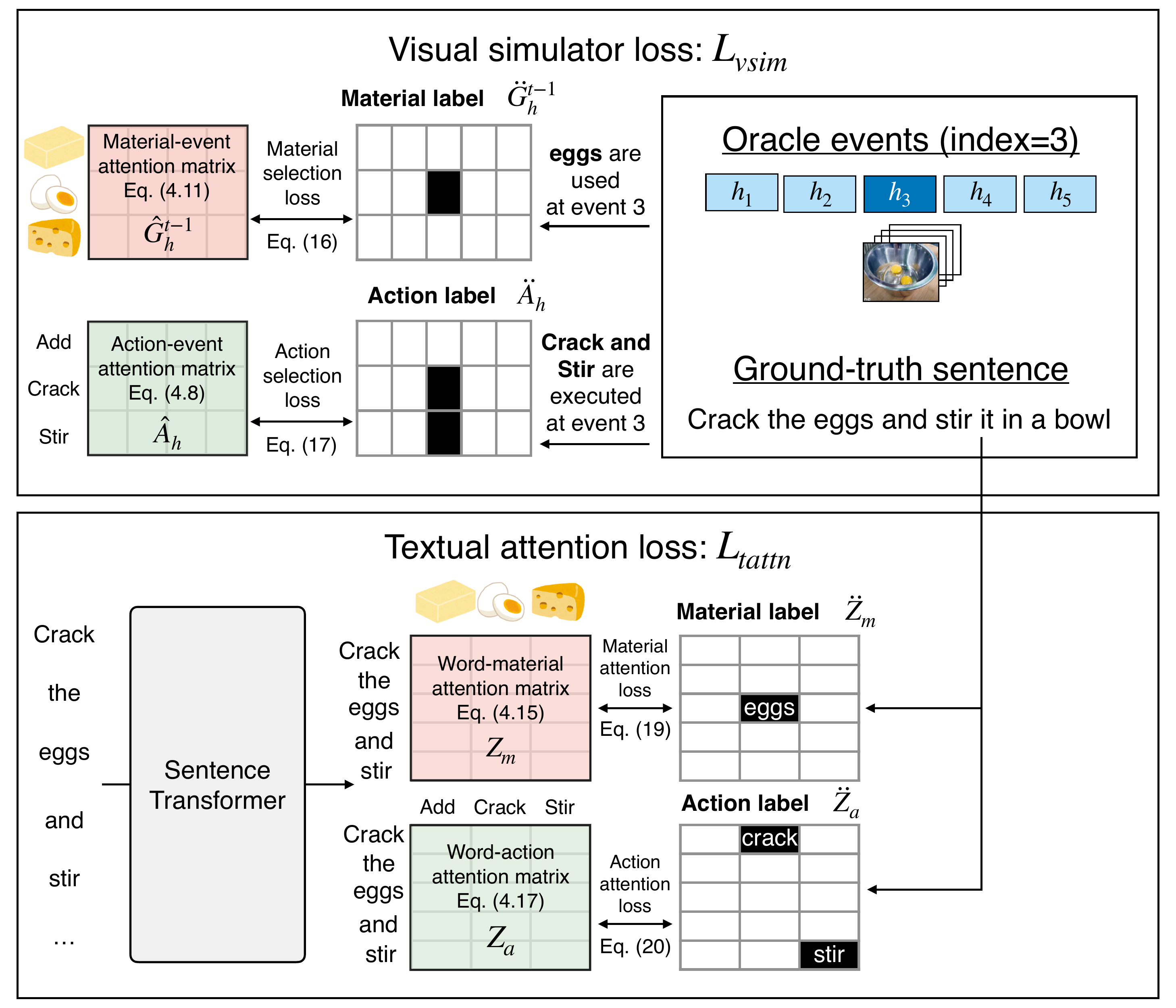}
  \caption{Overview of loss computation of visual selector loss and textual attention loss.}
  \label{fig:loss_computation}
\end{figure}

In addition to the losses described in \secref{subsec:loss_function}, we introduce the following two types of loss functions: (1) visual simulator loss $L_{vsim}$ and (2) textual attention loss $L_{tattn}$. \figref{fig:loss_computation} shows an overview of the loss computation for these two losses.

\textbf{(1) Visual simulator loss $L_{vsim}$} aims to train the visual simulator, summing up two losses: (1) ingredient selection loss $L_{is}$ and (2) action selection loss $L_{as}$.
Given the action- and ingredient-event attention matrices in \eqref{eq:adot} and \eqref{eq:gdot}, they are computed as the summed negative log-likelihood based on the ingredients/actions and events that constitute the recipe story.

To avoid costly human annotations, we compute the loss through distant supervision \cite{mintz2009acl} following our previous work.
For the ingredient selection loss, labels are obtained on whether the sentence corresponding ground-truth event contains ingredients and whether the events are oracle or not at each step.

For the action selection loss, labels are obtained whether the sentence has actions in the 384 actions defined by \cite{bosselut2018iclr} and whether the events are oracle or not at each step.
For example, in \figref{fig:loss_computation}, ``eggs'' at the event index of 3 is extracted as an ingredient label, and ``crack'' and ``stir'' are extracted at the same event index as an action label.
Mathematically, let $\ddot{A}_h, \ddot{G}_{h}^{t-1}$ be the action- and material-label binary matrices respectively, where each element represents $0$ or $1$. Based on $\hat{A}_h, \hat{G}_h^{t-1}, \ddot{A}_h, \ddot{G}_{h}^{t-1}$, the ingredient and action selection losses are computed as:
\begin{eqnarray}
    L_{is} &=& - \mathrm{SumMat}({\log({\hat{G}_h^{t-1} \odot \ddot{G}_{h}^{t-1}})}), \\
    L_{as} &=& - \mathrm{SumMat}({\log({\hat{A}_h \odot \ddot{A}_h)}}), \\
    L_{vsim} &=& L_{is} + L_{as},
\end{eqnarray}
where $\mathrm{SumMat(\cdot)}$ represents the sum of all of the elements of the matrix.

\textbf{(2) Textual attention loss $L_{tattn}$} aims to train the textual attention, consisting of two losses: (1) ingredient attention loss $L_{ia}$ and (2) action attention loss $L_{aa}$.
Given the word-ingredient/action attention matrices in \eqref{eq:word_gdot} and \eqref{eq:word_adot}, they are computed as the sum of the negative log-likelihood based on whether the $k$-th word is the same as the ingredients/actions.
Mathematically, let $\ddot{\Z}_a,\ddot{\Z}_m$ be the action- and material-word binary matrices, which represent whether the $k$-th word is the same as the actions/ingredients.
Given the word-ingredient/action attention matrices $\Z_m, \Z_a$ and $\ddot{\Z}_a,\ddot{\Z}_m$, they are computed as:
\begin{eqnarray}
    L_{ia} &=& - \mathrm{SumMat}({\log({\Z_m \odot \ddot{\Z}_m})}), \\
    L_{aa} &=& - \mathrm{SumMat}({\log({\Z_a \odot \ddot{\Z}_a})}), \\
    L_{tattn} &=& L_{ia} + L_{aa}.
\end{eqnarray}

\textbf{Total loss.}
We simply add these losses to the loss defined in \secref{subsec:loss_function} as follows:
\begin{equation}
    L_{extended} = L_{base} + L_{vsim} + L_{tattn}.
\end{equation}

\section{Experiments}
We use the YouCook2 dataset \cite{zhou2018aaai}, which consists of 2,000 cooking videos from 89 recipe categories.
All of the videos have 3--16 clips with a human-annotated start/end timestamp, and each clip is also annotated with an English sentence. The average length of videos is 5.27 minutes per video.
The original YouCook2 dataset does not contain ingredient annotations, thus we use the YouCook2-ingredient dataset \cite{nishimura2021acmmm}, which contains additional ingredient annotations. The dataset also contains 1,788 videos with events, sentences, and ingredients. We use the official split proposed by \cite{zhou2018aaai} for evaluation. Because the test set is not available online, we use the validation set for evaluation.
When training the proposed models, we first set the pre-trained weights of event encoders (e.g., MIL-NCE), word embedding (GloVe), and ingredient embedding (GloVe).
Then we freeze only the event encoders and finally train other modules.
Note that the model is trained from scratch except for these pre-trained modules, and the frozen event encoder accounts for 18\% parameters (9.6M/52.7M) of total parameters.
For our experiments, we use a single NVIDIA RTX A5000 GPU.

\textbf{Event encoder.}
We employ different two encoders.
\begin{itemize}
    \item \textbf{TSN} \cite{wang2019pami} converts the appearance and optical flow into frame-level representations, and then outputs the event-level representations by averaging them. For appearance, we use 2,048D feature vectors extracted from the ``Flatten-673'' layer in ResNet-200 \cite{he2016cvpr}. For the optical flow, 1,024D feature vectors were extracted from the ``global pool'' layer in BN-Inception \cite{ioffe2015icml}. Note that these models are pre-trained on only vision resources (e.g., ImageNet \cite{deng2009cvpr}).
    \item \textbf{MIL-NCE} \cite{miech2020cvpr} converts the events into representations. The model is pre-trained on Howto100M \cite{miech2019iccv}, which consists of automatically constructed 100M clip-narration pairs. We expect the MIL-NCE to yield better event representations than the TSN because this model is pre-trained on instructional vision-and-language resources.
\end{itemize}

\textbf{Data preprocessing.}
As in \cite{lei2020acl}, we truncated sequences longer than 100 for the clip and 20 for the sentence and set the maximum length of the clip sequence to 12.
Finally, we built the vocabulary based on words that occurred at least three times. The resulting vocabulary contained 991 words.

\textbf{Hyper-parameter settings.}
For both the encoder and decoder transformers, we set the hidden size to 768, the number of layers to two, and the number of attention heads to 12.
We train the model using the optimization method described in \cite{dalvin2019naacl,lei2020acl}; we use the Adam optimizer \cite{diederik2015iclr} with an initial learning rate of $0.0001$, $\beta_1=0.9$, and $\beta_2=0.999$.
The L2 weight decay is set to $0.01$, and the learning rate warmup is over the first five epochs.
We set the batch size to 16, and continue training at most 50 epochs using early stopping with SODA:CIDEr-D.
The early stopping is introduced from MART, which stops the model training based on CIDEr-D.

\textbf{Models.}
We test the proposed method by comparing it with four state-of-the-art dense video captioning models, as described below:
\begin{itemize}
    \item \textbf{Masked Transformer (MT)} \cite{zhou2018cvpr} is a transformer-based encoder-decoder DVC model that can be trained in an end-to-end manner by using a differentiable mask.
    \item \textbf{Event-centric Hierarchical Representation for DVC (ECHR)} \cite{wang2021tcsvt} is an event-oriented encoder-decoder architecture for DVC. The ECHR incorporates temporal and semantic relations into the output events for generating captions accurately.
    \item \textbf{SGR} \cite{deng2021cvpr} is a top-down DVC model consisting of four processes. The model (1) generates an overall paragraph from the input video, (2) grounds the sentences with events in the video, (3) refines captions based on the grounded events, and (4) refines events, referring to the refined captions.
    \item \textbf{PDVC} \cite{wang2021iccv} is the state-of-the-art DVC model. It detects $N$ events densely, re-ranks the top $K$ of them, and generates sentences for the re-ranked top $K$ events. Note that $K$ is the prediction target. This model is used in our preliminary experiments described in \secref{sec:oracle_analysis}.
\end{itemize}

\textbf{Ablations.}
We examine the impact of the components of the proposed method through ablation studies on the following variations:
\begin{itemize}
 \item \textbf{Base model (B)} is the model introduced in \secref{sec:proposed_method}.
 \item \textbf{B + Ingredient (BI)} incorporates the ingredient encoder into the base model.
 \item \textbf{BI + Visual simulator (BIV)} incorporates the visual simulator into the BI model.
 \item \textbf{BIV + Textual attention (BIVT)} additionally incorporates the textual attention module into the BIV model.
\end{itemize}

\subsection{Quantitative evaluation}

\begin{table}[t]
\centering
\caption{Quantitative evaluation for the baselines and proposed methods. We evaluate them on the word-overlap metrics and computational complexity. The bold and underlined scores are the best and second-best among the comparative methods. V and VI represents video only and video+ingredients, respectively.}
\scalebox{0.6}{
\begin{tabular}{c|c|c|c|c|c|c|ccc|ccc}
\hline
& Input modality & FLOPs & \#parameters & Training time & Inference time & Video feature & \multicolumn{3}{c}{dvc\_eval} & \multicolumn{3}{c}{SODA} \\ \hline
& & & & & & & BLEU4   & METEOR   & CIDEr-D  & METEOR & CIDEr-D & tIoU  \\ \hline
MT & V & - & - & - & - & TSN & 0.30 & 3.18 & 6.10 & - & - & - \\
ECHR & V & - & - & - & - & TSN & - & 3.82 & - & - & - & - \\
SGR  & V & - & - & - & - & TSN & - & - & - & 4.35 & - & - \\
PDVC (reported) & V & - & - & - & - & TSN & 0.80 & 4.72 & 22.71 & 4.42 & - & - \\
PDVC (reproduced) & V & 148.00G & 20.64M & 2.71h & 0.06s & TSN & 0.56 & 5.80 & 21.47 & 3.99 & 15.10 & 27.80 \\
PDVC (comparable) & V & 362.72G & 48.10M & 8.98h & 0.09s & TSN & 0.45 & 4.42 & 15.63 & 3.06 & 7.34 & 24.24 \\ \hline
Base (B) & V & 37.21G & 47.09M & 2.41h & 0.17s & MIL-NCE & 1.04 & 6.03 & 24.98 & 5.45 & 25.09 & 33.23 \\
Base + Ingredients (BI) & VI & 78.65G & 47.32M & 3.67h & 0.17s & MIL-NCE & 1.39 & 7.18 & 31.07 & 6.44 & 31.69 & \textbf{35.10} \\
BI + Visual simulator (BIV) & VI & 78.9G & 52.7M & 4.72h & 0.19s & MIL-NCE & \underline{1.40} & \underline{7.27} & \underline{32.67} & \underline{6.46} & \underline{32.95} & 34.13 \\
BIV + Textual attention (BIVT) & VI & 78.9G & 52.7M & 4.79h & 0.18s & MIL-NCE & \textbf{1.92} & \textbf{8.04} & \textbf{37.24} & \textbf{7.29} & \textbf{38.93} & \underline{35.06} \\ \hline
Oracle & V & - & - & - & - & TSN & 0.97 & 7.68 & 36.30 & 9.64 & 35.09 & 71.16 \\ \hline
\end{tabular}
}
\label{tab:word_overlap_metrics}
\end{table}

\label{sec:quantitative_results}
\subsubsection{Word-overlap evaluation}
\tabref{tab:word_overlap_metrics} demonstrates the results of the word-overlap evaluation on both dvc\_eval and SODA with BLEU, METEOR, and CIDEr-D.
We observe that the base model consistently outperforms state-of-the-art captioning models by a significant margin in both evaluations. In the ablation, the BIV model outperforms the BI model, and the BIVT model further improves the BIV model. This indicates that both the dot-product visual simulator and the textual attention module are effective for accurate recipe generation.
We also note that the BI model achieves the best performance on SODA:tIoU, but performs worse than BIV and BIVT on other word-overlap metrics. This indicates that the additional modules especially contribute to verbalizing the correct recipes, rather than selecting oracle events accurately.
\subsubsection{Discussion on the computational complexity}
In \tabref{tab:word_overlap_metrics}, we also demonstrate the computational complexity from two perspectives: (1) FLOPs and the number of parameters (\#parameters) and (2) training and inference speed.

\textbf{(1) FLOPs and \#parameters.}
To calculate \#parameters and FLOPs, we use the thop \footnote{\url{https://github.com/Lyken17/pytorch-OpCounter}}, which automatically computes FLOPs by forwarding the input data.
Note that both FLOPs and \#parameters include the frozen part of the model (i.e., event encoder).
\tabref{tab:word_overlap_metrics} indicates that \#parameters of PDVC are smaller than the proposed models, but its FLOPs are bigger. We investigate the reason by analyzing layer-wise FLOPs and discover that the primary bottleneck is the LSTM decoder, which cannot be computed in parallel and costs $O(n * d^2)$ as discussed in \cite{vaswani2017neurips}. To fill the gap of \#parameters between PDVC and our methods, we increase the number of Transformer layers of PDVC (from $2$ to $10$) and compare it with the proposed methods. We refer to it as ``PDVC (comparable)'' in \tabref{tab:word_overlap_metrics}.
However, PDVC (comparable) underperforms both the proposed methods and the original PDVC model across all metrics.
This may be primarily attributed to its excessive parameters, leading to overfitting of the training data.

\textbf{(2) Training and inference speed.}
We compute the total training time and video-wise inference speed. We confirm that both the total training time and inference speed of PDVC are much shorter than our methods.
This occurs because PDVC can predict events and sentences in parallel, but our methods estimate the current ones recurrently by remembering the previously predicted results, causing increased computational costs.

\subsection{Discussion on the number of predicted events}

\begin{table}[t]
\centering
\caption{Percentage of recipes that satisfy $|p - q| \leq \eta$, where $p, q, \eta$ represents the number of predicted events, ground-truth events, and a threshold, respectively. In this experiment, we change $\eta$ from $0$ to $3$.}
\scalebox{1.0}{
\begin{tabular}{ccccc}
\hline
\multicolumn{1}{c|}{$\eta$} & 0 & 1 & 2 & 3 \\ \hline
\multicolumn{1}{c|}{PDVC} & 14.4 & 40.0 & 63.0 & 76.4 \\ \hline
Model & & & & \\ \hline
\multicolumn{1}{c|}{B} & 18.6 & \textbf{52.1} & 71.7 & 83.4 \\
\multicolumn{1}{c|}{BI} & 18.8 & 51.6 & \underline{73.5} & \underline{87.3} \\
\multicolumn{1}{c|}{BIV} & \textbf{20.5} & \underline{51.8} & \textbf{74.2} & 86.2 \\
\multicolumn{1}{c|}{BIVT} & \underline{19.7} & \underline{51.8} & \underline{73.5} & \textbf{87.5} \\ \hline
\end{tabular}
}
\label{tab:event_pred_num}
\end{table}

\begin{figure}[t]
  \centering
  \includegraphics[width=0.8\linewidth]{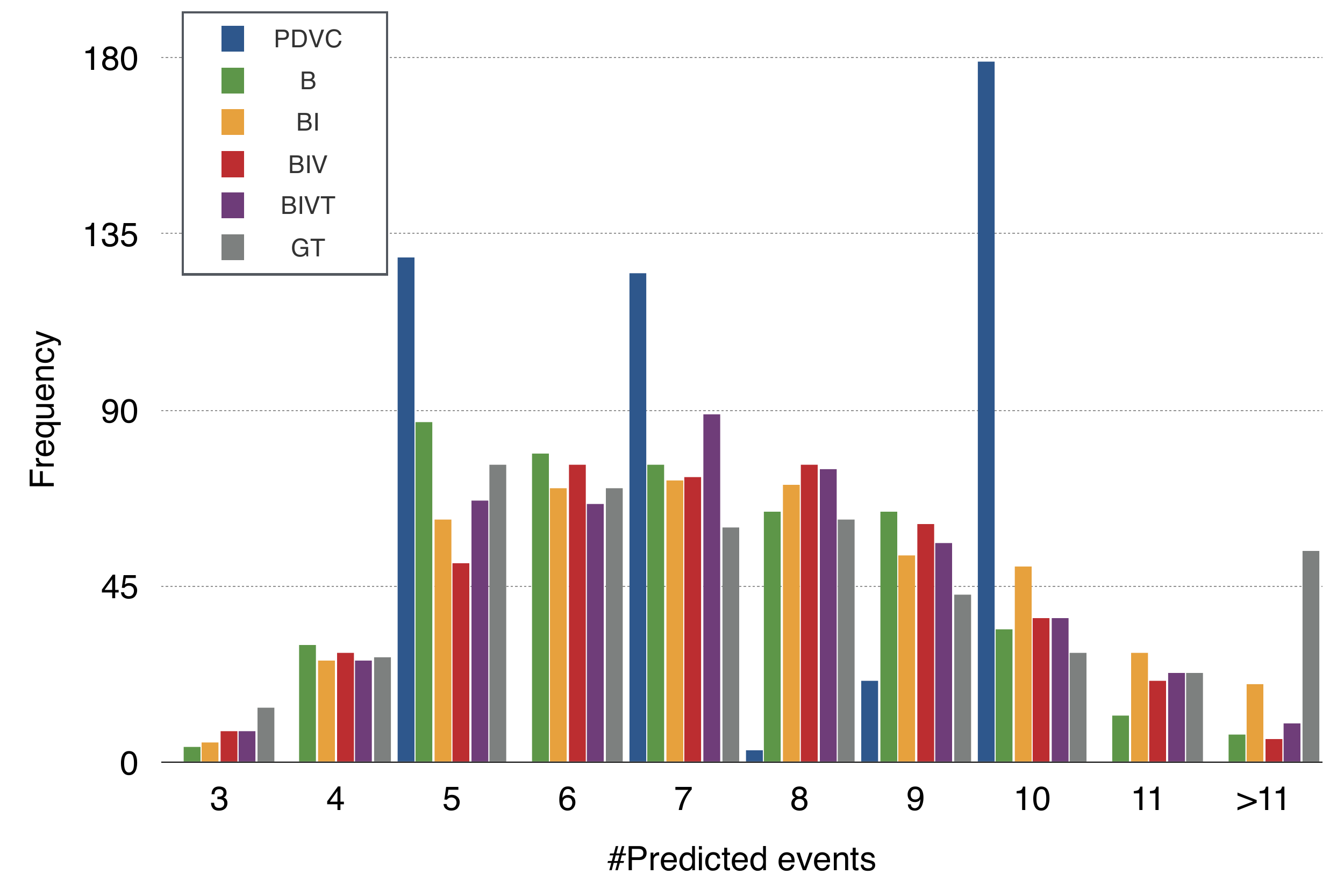}
  \caption{Histogram of the number of predicted events and ground truth.}
  \label{fig:predicted_event_histo}
\end{figure}

In addition to the word-overlap metrics, we discuss the generated recipes from the perspective of the number of predicted events. \tabref{tab:event_pred_num} shows the percentage of recipes that satisfy $|p - q| \leq \eta$, where $p, q, \eta$ represents the number of predicted events, ground truth, and a threshold, respectively. In this experiment, we change $\eta$ from $0$ to $3$. This result demonstrates that the proposed models consistently predict a more precise number of events than the PDVC.
\figref{fig:predicted_event_histo} shows the histogram of the number of the predicted events. While the PDVC outputs the specific number of events (i.e., 5, 7, 10), the histogram of the proposed method draws a similar curve to that of the ground truth.
In contrast, the ablation study does not show a clear consistency in the number of predicted events.
As discussed in \secref{sec:quantitative_results}, the additional modules can improve the word-event-based metrics but do not affect the period of events.

\subsection{Qualitative analysis}

\begin{figure}[t]
  \centering
  \includegraphics[width=0.8\linewidth]{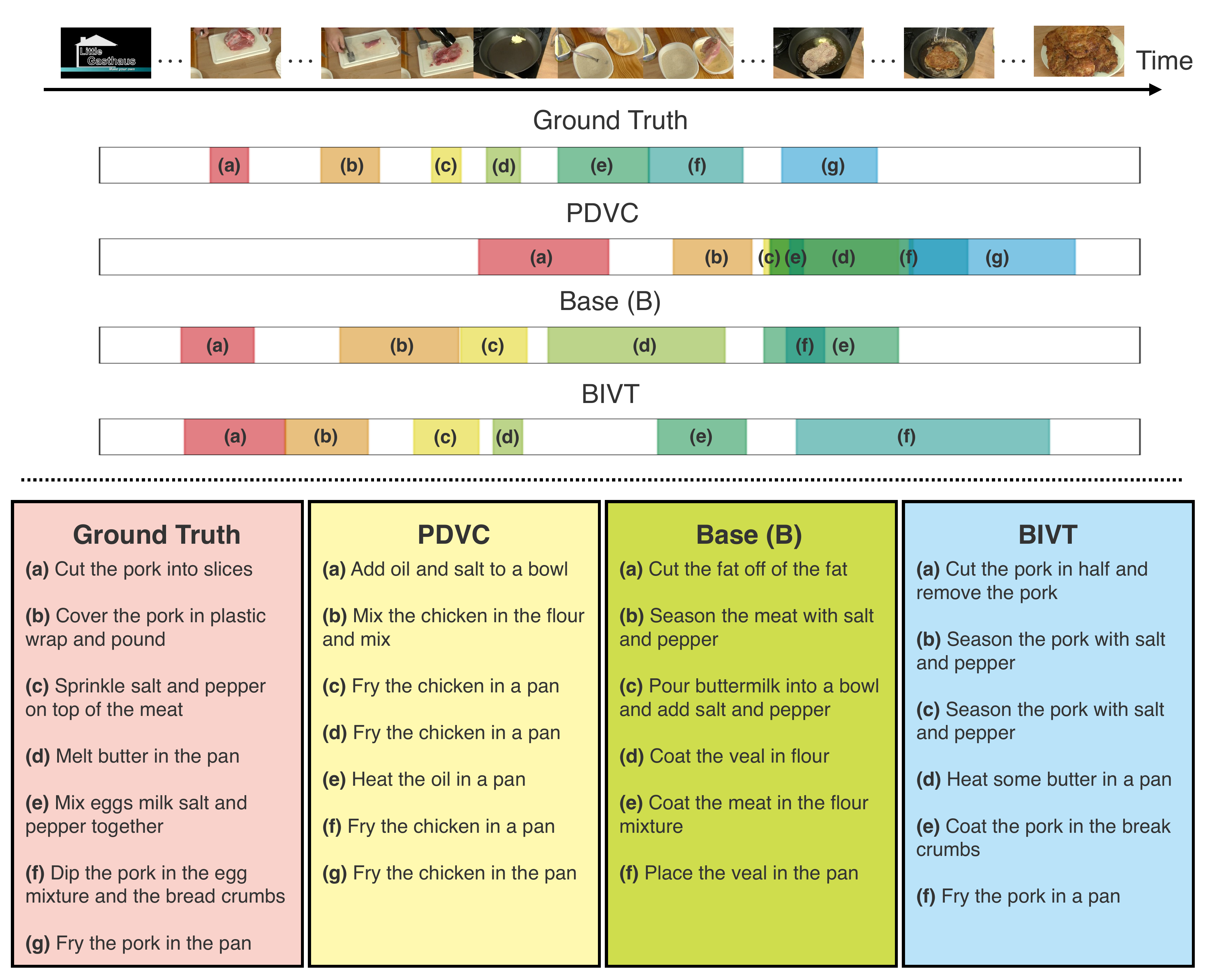}
  \caption{Examples of the generated recipes. We compare three models: PDVC, B, and BIVT with the ground truth.}
  \label{fig:generated_recipe}
\end{figure}

\figref{fig:generated_recipe} shows the predicted events and generated recipes from the PDVC, B, and BIVT models, in comparison to that of the ground truth.
In terms of the predicted events, PDVC outputs highly overlapped events.
The proposed method, on the other hand, predicts events in the correct order, with minimal overlap. This story-oriented sequential event prediction is an advantage of the proposed method.
In terms of the generated recipes, the PDVC repeatedly generates the same contents, ignoring the events (see (c) to (g)). The proposed methods suppress this problem. A comparison of B and BIVT reveals that the BIVT can generate sentences that are grounded with events (e.g., ``pork'' is accurately verbalized in (a) in BIVT, whereas ``fat'' is generated in (a) in B).

\subsection{Discussion on the detailed model settings}
Here, we discuss the detailed model settings from six perspectives: (1) loss ablation studies, (2) memory update strategies, (3) input modalities, (4) event encoders, (5) parameter sensitivity and the event candidates $N$, and (6) trade-off between the model's performance, computational overhead, and model complexity.
The results demonstrate that these parameters are important to succeed in our task.

\subsubsection{Loss ablation studies}

 \begin{table}[t]
\centering
\caption{Loss ablation studies. IS, AS, IA, and AA represent ingredient selection, action selection, ingredient attention, and action attention losses, respectively.}
\scalebox{0.8}{
\begin{tabular}{c|cccc|ccc|ccc}
\hline
& IS & AS & IA & AA & \multicolumn{3}{c|}{dvc\_eval} & \multicolumn{3}{c}{SODA} \\ \hline
& & & & & BLEU4 & METEOR & CIDEr-D & METEOR & CIDEr-D & tIoU  \\ \hline
(a) & & & & & 1.11 & 6.37 & 26.54 & 5.69 & 26.54 & 33.15 \\
(b) & \checkmark & & & & 1.28 & 7.31 & 31.59 & 6.46 & 33.10 & 33.78 \\
(c) & & \checkmark & & & 1.44 & 7.32 & 32.19 & 6.35 & 30.86 & 33.35 \\
(d) & \checkmark & \checkmark & & & 1.57 & 7.45 & 33.56 & 6.76 & 35.45 & 34.56 \\ \hline
(e) & \checkmark & \checkmark & \checkmark & & \underline{1.76} & \underline{7.79} & \underline{36.84} & \underline{7.12} & 36.96 & \underline{34.91} \\
(f) & \checkmark & \checkmark & & \checkmark & 1.57 & 7.70 & 34.58 & 6.93 & \underline{37.42} & 34.53 \\
(g) & \checkmark & \checkmark & \checkmark & \checkmark & \textbf{1.92} & \textbf{8.04} & \textbf{37.24} & \textbf{7.29} & \textbf{38.93} & \textbf{35.06} \\ \hline
\end{tabular}
}
\label{tab:loss_ablation}
\end{table}

\tabref{tab:loss_ablation} shows ablation studies on the loss function for the BIVT model. The experiment yields two insights.
First, all the losses are essential for training the model. This is confirmed by removing all of them from the full model (compare (a) and (g)).
Second, ingredient-based losses are more necessary than action-based ones.
Removing the ingredient selection loss has a more significant influence on the performance, compared to the action selection loss (compare (b) with (d) and (c) with (d)). This tendency is the same as the relationship between ingredient attention and action attention losses.

\subsubsection{Memory update strategies: separate or joint?}
\label{subsub:seperate_joint}

\begin{table}[t]
\centering
\caption{Comparison of memory update strategies: separate vs joint.}
\scalebox{0.8}{
\begin{tabular}{ccccccc}
\hline
\multicolumn{1}{c|}{} & \multicolumn{3}{c|}{dvc\_eval} & \multicolumn{3}{c}{SODA} \\ \hline
\multicolumn{1}{c|}{} & BLEU4 & METEOR & \multicolumn{1}{c|}{CIDEr-D} & METEOR & CIDEr-D & tIoU \\ \hline
B & & & & & & \\ \hline
\multicolumn{1}{c|}{Separate} & 0.92 & 5.67 & \multicolumn{1}{c|}{24.19} & 5.19 & 23.12 & \textbf{33.96} \\
\multicolumn{1}{c|}{Joint} & \textbf{1.04} & \textbf{6.03} & \multicolumn{1}{c|}{\textbf{24.98}} & \textbf{5.45} & \textbf{25.08} & 33.26 \\ \hline
BIV & & & & & & \\ \hline
\multicolumn{1}{c|}{Separate} & 1.39 & 7.24 & \multicolumn{1}{c|}{\textbf{33.02}} & \textbf{6.54} & \textbf{33.00} & 33.81          \\
\multicolumn{1}{c|}{Joint} & \textbf{1.40} & \textbf{7.27} & \multicolumn{1}{c|}{32.67} & 6.46 & 32.95 & \textbf{34.13} \\ \hline
BIVT & & & & & & \\ \hline
\multicolumn{1}{c|}{Separate} & 1.78 & 7.85 & \multicolumn{1}{c|}{36.65} & 6.92 & 36.84 & 33.85 \\
\multicolumn{1}{c|}{Joint} & \textbf{1.92} & \textbf{8.04} & \multicolumn{1}{c|}{\textbf{37.24}} & \textbf{7.29} & \textbf{38.93} & \textbf{35.06} \\ \hline
\end{tabular}
}
\label{tab:memory_update}
\end{table}

\tabref{tab:memory_update} shows a comparison of the memory update strategies.
While the separate memory update does not mix the memories in the event and sentence transformers, the joint approach proposed in \secref{sec:proposed_method} fuses these memories.
The results demonstrate that the joint approach outperforms the separate one on the base and BIVT settings, but on BIV, the separate approach performs the equivalent or better performance than the joint one.
Based on these observations, we confirm the effectiveness of the proposed memory mixing.
However, there is room for further research to improve the memory mixing method that is effective for all ablation models.

\subsubsection{Input modalities: video only or multimodal?}

\begin{table}[t]
\centering
\caption{Comparison of input modalities: video only and multimodal versions.}
\scalebox{0.8}{
\begin{tabular}{ccccccc}
\hline
\multicolumn{1}{c|}{} & \multicolumn{3}{c|}{dvc\_eval} & \multicolumn{3}{c}{SODA} \\ \hline
\multicolumn{1}{c|}{} & BLEU4 & METEOR & \multicolumn{1}{c|}{CIDEr-D} & METEOR & CIDEr-D & tIoU \\ \hline
B & & & & & & \\ \hline
\multicolumn{1}{c|}{Video only} & \textbf{1.04} & \textbf{6.03} & \multicolumn{1}{c|}{\textbf{24.98}} & \textbf{5.45} & \textbf{25.08} & \textbf{33.26} \\
\multicolumn{1}{c|}{Multimodal} & 0.44 & 3.91 & \multicolumn{1}{c|}{15.49} & 3.51 & 13.21 & 27.33 \\ \hline
BIV & & & & & & \\ \hline
\multicolumn{1}{c|}{Video only} & \textbf{1.40} & \textbf{7.27} & \multicolumn{1}{c|}{\textbf{32.67}} & \textbf{6.46} & \textbf{32.95} & \textbf{34.13} \\
\multicolumn{1}{c|}{Multimodal} & 0.60 & 4.68 & \multicolumn{1}{c|}{19.02} & 4.03 & 14.17 & 27.32 \\ \hline
BIVT & & & & & & \\ \hline
\multicolumn{1}{c|}{Video only} & \textbf{1.92} & \textbf{8.04} & \multicolumn{1}{c|}{\textbf{37.24}} & \textbf{7.29} & \textbf{38.93} & \textbf{35.06} \\
\multicolumn{1}{c|}{Multimodal} & 0.78 & 5.18 & \multicolumn{1}{c|}{21.00} & 4.66 & 19.51 & 28.94 \\ \hline
\end{tabular}
}
\label{tab:input_modality}
\end{table}

MIL-NCE has two branches of encoders: video and text encoders. In our experiments, we use only the video branch but can employ a text encoder by inputting the predicted events with generated captions of PDVC. We compare video only version and the multimodal version; in the multimodal version, we convert event timestamps and sentences into vectors using MIL-NCE branches, simply concatenate them, and input them into the models.
\tabref{tab:input_modality} shows a comparison of the input modalities, indicating that the video-only inputs achieve much better than the multimodal inputs. This occurs because the sentences generated by PDVC are semantically overlapped (see \figref{fig:generated_recipe}) and do not contribute to our recipe generation task.

\subsubsection{Event encoders}

\begin{table}[t]
\centering
\caption{Comparison of the model's performance when varying the event encoders: TSN and MIL-NCE. Note that unlike TSN, which is pre-trained on only vision resources, the MIL-NCE is pre-trained on instructional vision-and-language resource, Howto100M.}
\scalebox{0.8}{
\begin{tabular}{ccccccc}
\hline
\multicolumn{1}{c|}{} & \multicolumn{3}{c|}{dvc\_eval} & \multicolumn{3}{c}{SODA} \\ \hline
\multicolumn{1}{c|}{} & BLEU4 & METEOR & \multicolumn{1}{c|}{CIDEr-D} & METEOR & CIDEr-D & tIoU \\ \hline
B & & & & & & \\ \hline
\multicolumn{1}{c|}{TSN} & 0.36 & 4.24 & \multicolumn{1}{c|}{15.55} & 3.79 & 14.98 & 31.71 \\
\multicolumn{1}{c|}{MIL-NCE} & \textbf{1.04} & \textbf{6.03} & \multicolumn{1}{c|}{\textbf{24.98}} & \textbf{5.45} & \textbf{25.08} & \textbf{33.26} \\ \hline
BIV & & & & & & \\ \hline
\multicolumn{1}{c|}{TSN} & 0.52 & 4.93 & \multicolumn{1}{c|}{18.98} & 4.51 & 18.32 & 31.77 \\
\multicolumn{1}{c|}{MIL-NCE} & \textbf{1.40} & \textbf{7.27} & \multicolumn{1}{c|}{\textbf{32.67}} & \textbf{6.46} & \textbf{32.95} & \textbf{34.13} \\ \hline
BIVT & & & & & & \\ \hline
\multicolumn{1}{c|}{TSN} & 0.99 & 5.87 & \multicolumn{1}{c|}{23.60} & 5.26 & 22.84 & 32.36 \\
\multicolumn{1}{c|}{MIL-NCE} & \textbf{1.92} & \textbf{8.04} & \multicolumn{1}{c|}{\textbf{37.24}} & \textbf{7.29} & \textbf{38.93} & \textbf{35.06} \\ \hline
\end{tabular}
}
\label{tab:event_encoder}
\end{table}

\tabref{tab:event_encoder} shows the performance difference when changing the event encoders, indicating that the MIL-NCE proved significantly superior to the TSN in all of the settings. This occurs because the MIL-NCE is pre-trained on the vision-and-language resource, Howto100M, which captures the fine-grained event-level semantics of cooking procedures. We conclude that pre-training on an appropriate resource is essential to effective performance on our task.

\subsubsection{Parameter sensitivity and the event candidates $N$}

\begin{table}[t]
\centering
\caption{Comparison of the model's performance when varying the number of the event candidates $N$.}
\scalebox{0.8}{
\begin{tabular}{ccccccc}
\hline
\multicolumn{1}{c|}{}      & \multicolumn{3}{c|}{dvc\_eval}                & \multicolumn{3}{c}{SODA} \\ \hline
\multicolumn{1}{c|}{}      & BLEU4 & METEOR & \multicolumn{1}{c|}{CIDEr-D} & METEOR & CIDEr-D & tIoU  \\ \hline
B                          &       &        &                              &        &         &       \\ \hline
\multicolumn{1}{c|}{N=25}  & \textbf{1.27}  & \textbf{6.49}   & \multicolumn{1}{c|}{\textbf{27.84}}   & \textbf{6.19}   & \textbf{29.34}   & \textbf{35.26} \\
\multicolumn{1}{c|}{N=50}  & 0.98  & \underline{6.42}   & \multicolumn{1}{c|}{\underline{27.12}}   & \underline{5.89}   & \underline{26.34}   & 33.79 \\
\multicolumn{1}{c|}{N=100} & \underline{1.04}  & 6.03   & \multicolumn{1}{c|}{24.98}   & 5.45   & 25.08   & 33.26 \\
\multicolumn{1}{c|}{N=200} & 0.93  & 6.16   & \multicolumn{1}{c|}{25.93}   & 5.52   & 26.13   & \underline{33.93} \\ \hline
BIV                        &       &        &                              &        &         &       \\ \hline
\multicolumn{1}{c|}{N=25}  & \textbf{1.71}  & \underline{7.50}   & \multicolumn{1}{c|}{\textbf{34.18}}   & \textbf{7.02}   & \textbf{36.39}   & \textbf{36.13} \\
\multicolumn{1}{c|}{N=50}  & \underline{1.51}  & \textbf{7.52}   & \multicolumn{1}{c|}{\underline{33.37}}   & \underline{6.86}   & \underline{34.24}   & \underline{35.60} \\
\multicolumn{1}{c|}{N=100} & 1.40  & 7.27   & \multicolumn{1}{c|}{32.67}   & 6.46   & 32.95   & 34.13 \\
\multicolumn{1}{c|}{N=200} & 1.23  & 6.81   & \multicolumn{1}{c|}{29.51}   & 5.95   & 28.86   & 33.15 \\ \hline
BIVT                       &       &        &                              &        &         &       \\ \hline
\multicolumn{1}{c|}{N=25}  & \underline{1.87}  & 7.95   & \multicolumn{1}{c|}{36.12}   & \textbf{7.51}   & \textbf{39.06}   & \textbf{36.74} \\
\multicolumn{1}{c|}{N=50}  & 1.81  & \underline{7.99}   & \multicolumn{1}{c|}{\underline{36.48}} & \underline{7.36}   & 38.49   & \underline{36.38} \\
\multicolumn{1}{c|}{N=100} & \textbf{1.92}  & \textbf{8.04}   & \multicolumn{1}{c|}{\textbf{37.24}}   & 7.29   & \underline{38.93}   & 35.06 \\
\multicolumn{1}{c|}{N=200} & 1.78  & 7.66   & \multicolumn{1}{c|}{35.21}   & 6.77   & 35.64   & 33.48 \\ \hline
\end{tabular}
}
\label{tab:change_of_N}
\end{table}

The PDVC allows users to select the number of candidates $N$ when training the model. In the original study on the PDVC, $N$ is set to be $100$ on the YouCook2 dataset. In this experiment, we change the parameter $N$ to be $25, 50, 100$, and $200$ to investigate the parameter sensitivity of the model.
\tabref{tab:change_of_N} shows the performance change of the proposed method in different model settings: B, BIV, and BIVT.
The results demonstrate that $N=25$ consistently performs better than other settings on SODA, but is not always best on dvc\_eval.
For example, $N=100$ on the BIVT model achieves better than $N=25$ on dvc\_eval.
Therefore, we could not find the optimal $N$ that achieves the consistently best performance on all of the ablation methods but discovered that the lower $N$ performs better, especially on the SODA scores.
In addition, we observe that increasing $N$ extremely degrades the model's verbalization ability (e.g., $6.19 \rightarrow 5.52$ on SODA:METEOR by changing $N=25 \rightarrow N=200$ on the base model in \tabref{tab:change_of_N}).
Although a higher $N$ makes the maximum tIoU larger (shown in \secref{sec:oracle_analysis}), the ratio of events that are not oracle but highly overlapped with the ground truth increases. This prevents the model from selecting oracle events precisely and causes it to overfit the training set.

\subsubsection{Trade-off between the model's performance, computational overhead, and model complexity.}

\begin{figure}[t]
  \centering
  \includegraphics[width=0.8\linewidth]{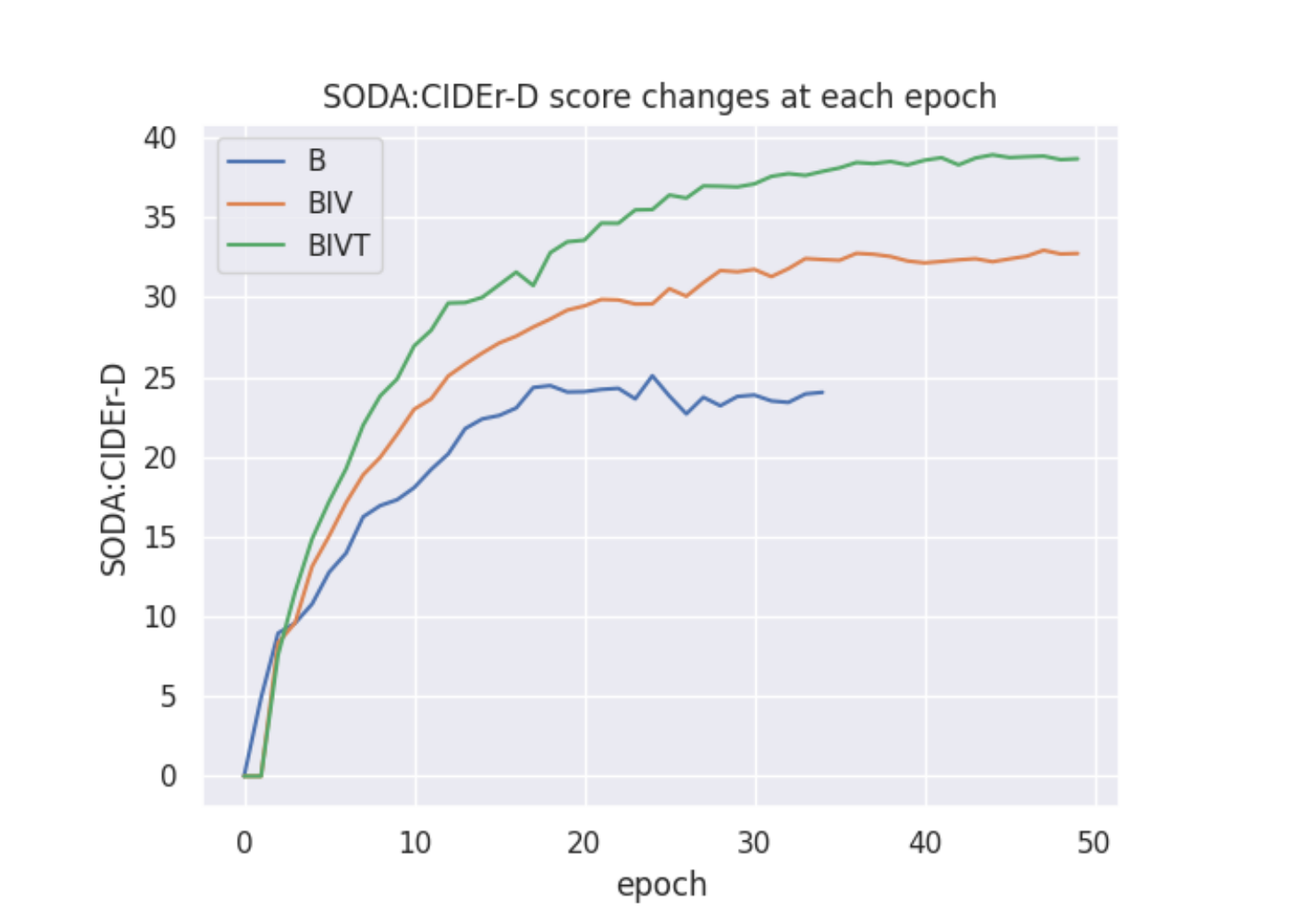}
  \caption{SODA:CIDEr-D changes at each epoch. Note that training the base model is stopped at epoch 35 due to early stopping.}
  \label{fig:epoch_cider}
\end{figure}

\begin{table}[t]
\centering
\caption{SODA score changes when changing the number of transformer layers in both event selector and sentence generator.}
\scalebox{1.0}{
\begin{tabular}{cccc|ccc|ccc}
\hline
& \multicolumn{3}{c|}{SODA: METEOR} & \multicolumn{3}{c|}{SODA:CIDEr-D}    & \multicolumn{3}{c}{SODA:tIoU} \\ \hline
\multicolumn{1}{c|}{} & 1 & 2 & 4 & 1 & 2 & 4 & 1 & 2 & 4 \\ \hline
\multicolumn{1}{c|}{B}    & \underline{5.39} & \textbf{5.45} & 5.25 & \textbf{26.26} & 25.09 & \underline{25.52} & \underline{32.52} & \textbf{33.23} & 32.06 \\
\multicolumn{1}{c|}{BIV}  & \textbf{6.47} & \underline{6.46} & 6.33 & \underline{32.59} & \textbf{32.95} & 31.44 & \underline{34.17} & 34.13 & \textbf{34.87} \\
\multicolumn{1}{c|}{BIVT} & 6.94 & \textbf{7.29} & \underline{7.11} & 36.96 & \textbf{38.93} & \underline{38.30} & 33.59 & \textbf{35.06} & \underline{34.90} \\ \hline
\end{tabular}
}
\label{tab:num_layers}
\end{table}

To discuss the trade-off between the model's performance and computational overhead, we demonstrate the SODA:CIDEr-D changes at each epoch in \figref{fig:epoch_cider}. This indicates that $30-40$ epochs are necessary for training the base model and $50$ are necessary for BIV and BIVT, concluding that 2.4 and 4.8 hours are necessary for achieving the best performance on B and BIV/BIVT, respectively (see \tabref{tab:word_overlap_metrics}).
Then, to discuss the trade-off between the model's performance and model complexity, we investigate the SODA score changes when changing the number of transformer layers $l=1,2,4$ in \tabref{tab:num_layers}.
The reason to investigate $l$ is that transformer layers are dominant in terms of the number of parameters of the models and increasing $l$ gains the model's complexity and computational costs.
The results demonstrate that (1) in general, $l=2$ generally performs the best on all of the settings and (2) $l=1$ works well on B and BIV while $l=4$ achieves better than $l=1$ on BIVT.
Therefore, we conclude that $l$ is an essential parameter to be tuned, but $l=2$ shows a good balance between performance and model complexity.

\section{Conclusion}
In this paper, we tackled recipe generation from unsegmented cooking videos, a task that requires agents to (1) extract key events that are essential to dish completion and (2) generate sentences for the extracted events.
We first analyzed the state-of-the-art DVC models and set our goal to obtain correct recipes by selecting oracle events from the output events of the DVC model and re-generating sentences for them.
To achieve this, we proposed a transformer-based multimodal recurrent learning model, which consists of the event selector and sentence generator.
The event selector selects oracle events from the event candidate in the correct order, and the sentence generator outputs a recipe grounded in the events.
Both modules have memory vectors to remember the history of previous predictions to estimate the next step.
The proposed memory mixing approach efficiently combines them, effectively sharing the previous predictions between the event selector and sentence generator.
To generate more accurate recipes, we also proposed an extended model by introducing two additional modules: dot-product visual simulator and textual attention module.
In experiments, we confirmed that the proposed methods achieve story awareness.
The base model outperforms the state-of-the-art DVC models and the extended model boosts the model's performance.
In addition, we showed that the proposed models can select the correct number of events, as with the ground-truth events.
The qualitative evaluation revealed that the proposed approaches can select events in the correct order and generate recipes grounded in the video content. Finally, we discussed the detailed experimental settings for optimal recipe generation.

\section*{Acknowledgement}
This work was supported by JSPS KAKENHI Grant Number JP21J20250, 21H04910, 20H04210, and JST-Mirai Program Grant Number JPMJMI21G2.

\bibliographystyle{ACM-Reference-Format}
\bibliography{sample-base}

\appendix

\end{document}